\documentclass[twocolumn,pra,showpacs,superscriptaddress,amssymb,amsmath]{revtex4}
\usepackage{graphicx}
\usepackage{epstopdf}
\usepackage{hyperref}
\hypersetup{%
  pdfpagemode=UseNone, %FullScreen,n
  pdfstartpage=1,
  pdfstartview=FitH,
  pdfmenubar=true,
  pdftoolbar=true,
  colorlinks = true,
  linkcolor=blue,
  citecolor=blue,
  bookmarksopen=false
}

\begin{document}

\title{Asymptotic model for shape resonance control of diatomics
  by intense non-resonant light: Universality in the single-channel
  approximation} 

\author{Anne Crubellier}
\email{anne.crubellier@u-psud.fr}
\affiliation{Laboratoire Aim\'e Cotton, CNRS, Universit\'e Paris-Sud 11, 
  ENS Cachan, B\^atiment 505, 91405  Orsay Cedex, France} 

\author{Rosario Gonz\'alez-F\'erez}
\email{rogonzal@ugr.es}
\affiliation{Instituto Carlos I de F\'isica
    Te\'orica y Computacional and Departamento de F\'isica At\'omica,
    Molecular y Nuclear, Universidad de Granada, 18071 Granada,
    Spain}
\affiliation{The Hamburg Center for Ultrafast Imaging, University of Hamburg,
      Luruper Chaussee 149, 22761 Hamburg, Germany}
\author{Christiane P. Koch}
\email{christiane.koch@uni-kassel.de}
\affiliation{Theoretische Physik,  Universit\"at Kassel,
  Heinrich-Plett-Str. 40, 34132 Kassel, Germany}

\author{Eliane Luc-Koenig}
\email{eliane.luc@u-psud.fr}
\affiliation{Laboratoire Aim\'e Cotton, CNRS, Universit\'e Paris-Sud 11, 
  ENS Cachan, B\^atiment 505, 91405  Orsay Cedex, France}  

\date{\today}
\begin{abstract}
Non-resonant light interacting with diatomics via the polarizability
anisotropy couples different rotational states and may lead to strong
hybridization of the motion. The modification of
shape resonances and low-energy scattering states due to this
interaction can be fully captured by an asymptotic model, based on the
long-range properties of the scattering [Crubellier et
al. arXiv:1412.0569]. Remarkably, the properties of the field-dressed
shape resonances in this asymptotic multi-channel description are found
to be approximately linear in the field intensity up to fairly large 
intensity. This suggests a perturbative single-channel approach to 
be sufficient to study the control of such resonances by the
non-resonant field.  
The multi-channel results furthermore indicate the dependence on field 
intensity to present, at least approximately, universal
characteristics. Here we combine the nodal line technique to solve the
asymptotic Schr\"odinger equation with perturbation theory. Comparing
our single channel results to those obtained with 
the full interaction potential, we find nodal lines depending 
only on the field-free scattering length of the diatom to yield an
approximate but universal description of the field-dressed molecule,
confirming universal behavior.
\end{abstract}

\pacs{34.50.Cx,34.50.Rk}
% 34.50.Cx Elastic; ultracold collisions 
% 34.50.Rk Laser-modified scattering and reactions 
\maketitle

\section{Introduction}
\label{sec:intro}

Quantum collisions at low energy depend on the long-range properties of the
interaction between the particles only  and therefore exhibit universal
behavior. Since, at long range, 
the dependence of the interaction on interparticle
distance has a power-law form  and it is often
sufficient to account only for the highest order term of the
interaction, low-energy 
collisions can be well described by simple models with very few
free parameters. This is at the core of multi-channel quantum defect
theory~\cite{SeatonJPB77,GreenePRA79,MiesJCP84,MiesJCP84b,%
GaoPRA98,GaoPRA01,JachymskiPRL13,JachymskiPRA14}. Universality becomes
particularly transparent when introducing units which absorb all
molecule-specific parameters~\cite{CrubellierJPB06}. The corresponding
Schr\"odinger equation can be solved by the so-called nodal line
technique~\cite{CrubellierEPJD99,PasquiouPRA10,VanhaeckeEPJD04}. It
consists in accounting for all short-range physics by the choice of 
position of the nodes of the scattering wavefunction at intermediate
interparticle distances. This formalism has been extended to shape
resonances~\cite{LondonoPRA10} and to the control of shape resonances
by non-resonant light which couples the different partial
$\ell$-waves via the polarizability
anisotropy~\cite{crubellierNJP14}. In particular, we have 
previously shown that an intensity-dependent nodal line is sufficient
to account for the effect of the coupling to the
non-resonant light at short-range. An asymptotic multi-channel model
can thus predict the resonance structure, energy, width and
hybridization as a function of non-resonant light intensity. 
This is important since non-resonant light control has been suggested
to enhance photoassociation rates~\cite{Aganoglu11,GonzalezPRA12},
modify Feshbach resonances~\cite{TomzaPRL14} and manipulate molecular 
levels~\cite{GonzalezPRA12,TomzaMolPhys13}.  
 
While a multi-channel treatment is essential to describe the strong 
hybridization of the rovibrational motion due to the coupling with the
non-resonant light~\cite{Aganoglu11,GonzalezPRA12}, the position and
width of the resonance are found to vary linearly with field intensity
up to fairly large intensities~\cite{crubellierNJP14}. When treating
the interaction with the non-resonant light as a perturbation and
truncating the perturbation expansion at the first order, resonance
position and width are determined by the field-free
wavefunctions. The field-free wavefunctions reside in a single partial
wave (channel) and, within the asymptotic approximation, depend on only
one parameter -- the background scattering length. 
This indicates universality of the intensity dependence of resonance
positions and widths in non-resonant light control. It furthermore 
suggests that a single-channel approach should be sufficient to study
non-resonant light control at moderate intensities. 

Here we combine the asymptotic model for shape resonance control with
non-resonant light developed in a preceding
paper~\cite{crubellierNJP14} with perturbation theory to explore the
universality of the resonance's intensity dependence. This allows us
to recast the multi-channel approach of Ref.~\cite{crubellierNJP14} in
a single channel approximation. We compare the perturbative results 
to those obtained in Ref.~\cite{crubellierNJP14} with 
a multi-channel description, by solving the 
Schr\"odinger equation for the diatom interacting with  non-resonant light
both with the full potential
and the asymptotic approximation. 

The paper is organized as follows: We briefly review
the asymptotic model for a diatomic interacting with non-resonant
light via the polarizability anisotropy
introduced in Ref.~\cite{crubellierNJP14}, hereafter referred to as
paper~I, in Sec.~\ref{sec:method}. We summarize the behavior of shape
resonances in non-resonant light observed by solving the multi-channel 
Schr\"odinger equation and present an approximate general 
law for describing the intensity dependence when analyzing the
resonances in reduced units of length and energy in
Sec.~\ref{sec:heuristic}. We then show in Sec.~\ref{sec:pert} 
how perturbation theory, either using energy-normalized continuum states 
(Sec.~\ref{subsec:continuum}) or a discretized continuum 
(Sec.~\ref{subsec:rosario}), is applied to determine the slopes of the
intensity dependence of position and width of the field-dressed shape
resonances at vanishing intensity. This allows us to explain the rule
observed for the specific example of strontium dimers considered in
paper~I.  
Then, section Sec.~\ref{sec:single-zero} describes systematic
single-channel calculations that   
allow for predicting 
the position and width of shape resonances without a non-resonant field.  
These results are used in Sec~\ref{sec:single-low-int} to deduce the
slope of the energy shifts of a shape resonance in the limit of
vanishing intensity for any angular momentum $\ell$, in any diatomic system.
We conclude in Sec.~\ref{sec:conclusion}. 

\section{Asymptotic model for a diatom in a 
  non-resonant optical field} 
\label{sec:method}

In this section, we summarize the theoretical framework for 
studying the interaction 
of a diatom with non-resonant light. A detailed derivation of the
asymptotic model is found in Ref.~\cite{crubellierNJP14}.
The Hamiltonian of an atom pair, with reduced 
mass $\mu$, interacting with a non-resonant laser field of intensity
$I$, linearly polarized along the space-fixed $Z$ axis,  
is written in the molecule-fixed frame as 
\begin{equation}
  \label{eq:2D_Hamil}
  H =   T_R+\frac{{\mathbf{L}}^2}{2\mu  R^2}+V_g(R)
  -\frac{2\pi I}{c}\left(\Delta\alpha(R)\cos^2\theta+\alpha_\perp(R)\right)\,.
\end{equation}
Here, $R$ denotes interatomic separation and $V_g(R)$ the interaction
potential in the electronic ground state. 
$T_R$ and $\mathbf{L}^2/2\mu R^2$ are the vibrational and 
rotational kinetic energies. In the last term of Eq.~\eqref{eq:2D_Hamil},
$c$ is the  speed of light and $\theta$  the polar angle between
the molecular axis and the laser polarization. The molecular
polarizability tensor is characterized by its  perpendicular and
parallel components with respect to the molecular axis $\alpha_\perp(R)$ and
$\alpha_\parallel(R)$, and the polarizability anisotropy is 
$\Delta\alpha(R)=\alpha_\parallel(R)-\alpha_\perp(R)$.
The non-resonant field introduces a mixing of different partial waves
$\ell$ of the same parity (channels), whereas the magnetic quantum
number $m$ is conserved. The corresponding multi-channel 
Schr\"odinger equation 
can be solved numerically as described in Ref.~\cite{GonzalezPRA12}. 

At sufficiently large distance, $R>R_{asym}=\sqrt{C_8/C_6}$, the
potential reduces to the asymptotic  
van der Waals interaction $V_g(R)\approx - C_6/R^6$
($C_n$ are the coefficients of the multipolar expansion).
For $R>R_d=(4\alpha_1 \alpha_2)^{1/6}$, where $\alpha_1$ $\alpha_2$ 
denote the polarizabilities of the two atoms, the interaction with the 
field reduces to
\begin{equation}
  \label{eq:Hfield}
  H_{int} = -\frac{2\pi I}{c}\left[(\alpha_1 +\alpha_2)
    +\frac{2\alpha_1 \alpha_2}{R^3}(3\cos^2\theta - 1)\right]\,. 
\end{equation}
Introducing a dimensionless reduced length $x$, $R  =  \sigma x$, 
reduced energy $e$, $E - E_0 =  \epsilon \,e$
(defined with respect to the field shifted dissociation limit 
$E_0=-\frac{2\pi }{c}(\alpha_1+\alpha_2)I$), and 
reduced  laser field intensity $i$,
$I = \beta~i$, and replacing all $R$-dependent terms by their leading
order contributions, an asymptotic Schr\"odinger equation is obtained, 
\begin{equation}
  \label{eq:asy}
  \left[-\frac{d^2}{dx^2} - \frac{1}{x^6} + \frac{\mathbf{L}^2}{x^2}
    - i
    \frac{\cos^2\theta -1/3}{x^3}
    - e
  \right] f (x,\theta) = 0\,.
\end{equation}
The unit conversion factors are given by 
\begin{equation}
\label{eq:scaling}
\sigma  =  \left(\frac{2\mu C_6}{\hbar^2}\right)^{1/4}\,,\;\;\; %\nonumber
\epsilon =  \frac{\hbar^2}{2\mu\sigma^2}\,, \;\;\; %\nonumber
\beta    
=\frac{c\sigma^3\epsilon}{12\pi\alpha_1 \alpha_2}\,,
\end{equation}
and the conversion factor for time is obtained from that of energy, 
$\tau=\hbar/\epsilon$. 
For each partial wave $\ell$, the wave function is expanded in terms
of Legendre polynomials $P_{\ell}(\cos\theta)$
\begin{equation}
\label{eq:fct}
 f_\ell(x,\theta) =  y_{\ell}(x) P_{\ell}(\cos\theta)\,, 
\end{equation}
and Eq.~\eqref{eq:asy} is solved in the asymptotic $x$-domain,
imposing to the radial 
function physical boundary condition at long range and to have a node at 
$x_{0\ell}>x_{asym}=R_{asym}/\sigma>x_d=R_d/\sigma$, a position
separating inner zone and asymptotic outer region, see
Ref.~\cite{crubellierNJP14} for details. 
The effects of potential $V_g(R)$, centrifugal 
energy  and polarizability in the inner zone 
can satisfactorily be accounted for by introducing   
energy-, angular-momentum- and intensity-dependent nodal lines
$x_{0\ell}$~\cite{LondonoPRA10,crubellierNJP14}, 
\begin{equation}
\label{eq:nodal_line_i}
{x}_{0\ell}=x_{00}+A~e+B\ell(\ell+1)+Ci\,.
\end{equation}
When the constants $A$, $B$ and $C$ can be deduced from exact
calculations using the Hamiltonian~\eqref{eq:2D_Hamil} or from
experiment, the nodal line technique applied to the asymptotic model
fully reproduces the results obtained
with~\eqref{eq:2D_Hamil}, cf. Fig.~2 of Ref.~\cite{crubellierNJP14}.  
When this is not possible, analytical values for $A^G$, 
$B^G$~\cite{LondonoPRA10} and $C^G$~\cite{crubellierNJP14} 
which depend on $x_{00}$, i.e., on the $s$-wave scattering length, 
allow for an approximate, universal description of  shape
resonances, very  
similar to the asymptotic model developed by Gao for field-free
resonanes~\cite{GaoJPB03}.
%
%---------------------------------------------------------------------------
\section{Heuristic scaling rule}
\label{sec:heuristic}
%--------------------------------------------------------------------------- 
We present and discuss in this section results obtained by solving the
Schr\"odinger equation with the exact Hamiltonian~\eqref{eq:2D_Hamil}.
Specifically, we consider the shape resonances with field-free 
$\ell =4,~8,~12,~16$ for 
$^{88}$Sr$_2$~\cite{GonzalezPRA12}, $\ell=5,~9$ for $^{133}$Cs$_2$, 
and $\ell=2$ for $^{87}$Rb$_2$. In paper I~\cite{crubellierNJP14},
a linear dependence of the resonance
position  vs field intensity was found 
for the two considered isotopes $^{88}$Sr$_2$ and $^{86}$Sr$^{88}$Sr 
up to very large values of the intensity~\cite{crubellierNJP14}. 
The use of reduced units (see \autoref{tab:redunits}) allows us to
extract from these results a general trend for the intensity
dependence of the resonance position, valid also for different  
$\ell$ values and different species. 
%****************************** tab 1 ****************************************
\begin{table}[tb]
  \centering
  \begin{tabular}{|l|c|c|c|c|}
    \hline 
    &$\sigma$ [a$_0$]& $\epsilon$ [$\mu$K] & 
    $\beta$ [$\,$GW/cm$^{2}$]& $\tau$ [ns]\\\hline
    $^{88}$Sr$_2$ I &  151.053 &  86.3653 & 
    0.635782 &   88.4409 \\ \hline
    $^{86}$Sr$^{88}$Sr II &  150.617 &  87.876 & 
    0.641319 &  86.9204 \\ \hline
    $^{133}$Cs$_2$  & 201.843 & 31.99 &  0.12048 &  238.7 \\ \hline
    $^{87}$Rb$_2$ & 165.250 & 72.99 & 0.25844 & 104.5 \\ \hline
  \end{tabular}
  \caption{Scaling factors, Eq.~\eqref{eq:scaling}, for 
    $^{88}$Sr$_2$ (case I) and $^{86}$Sr$^{88}$Sr (case II), obtained
    for $C_6=3246.97\,$a$_0^6$ and $\alpha_0=186.25\,$a$_0^3$, and for
    $^{133}$Cs$_2$ ($C_6=6851.0\,$a$_0^6$, $\alpha_0=402.20\,$a$_0^3$) and
    $^{87}$Rb$_2$ ($C_6=4707.0\,$a$_0^6$, $\alpha_0=309.98\,$a$_0^3$).  
  } 
  \label{tab:redunits}
\end{table}

\begin{figure}[tb]
  \centering
  \includegraphics[width=0.95\linewidth]{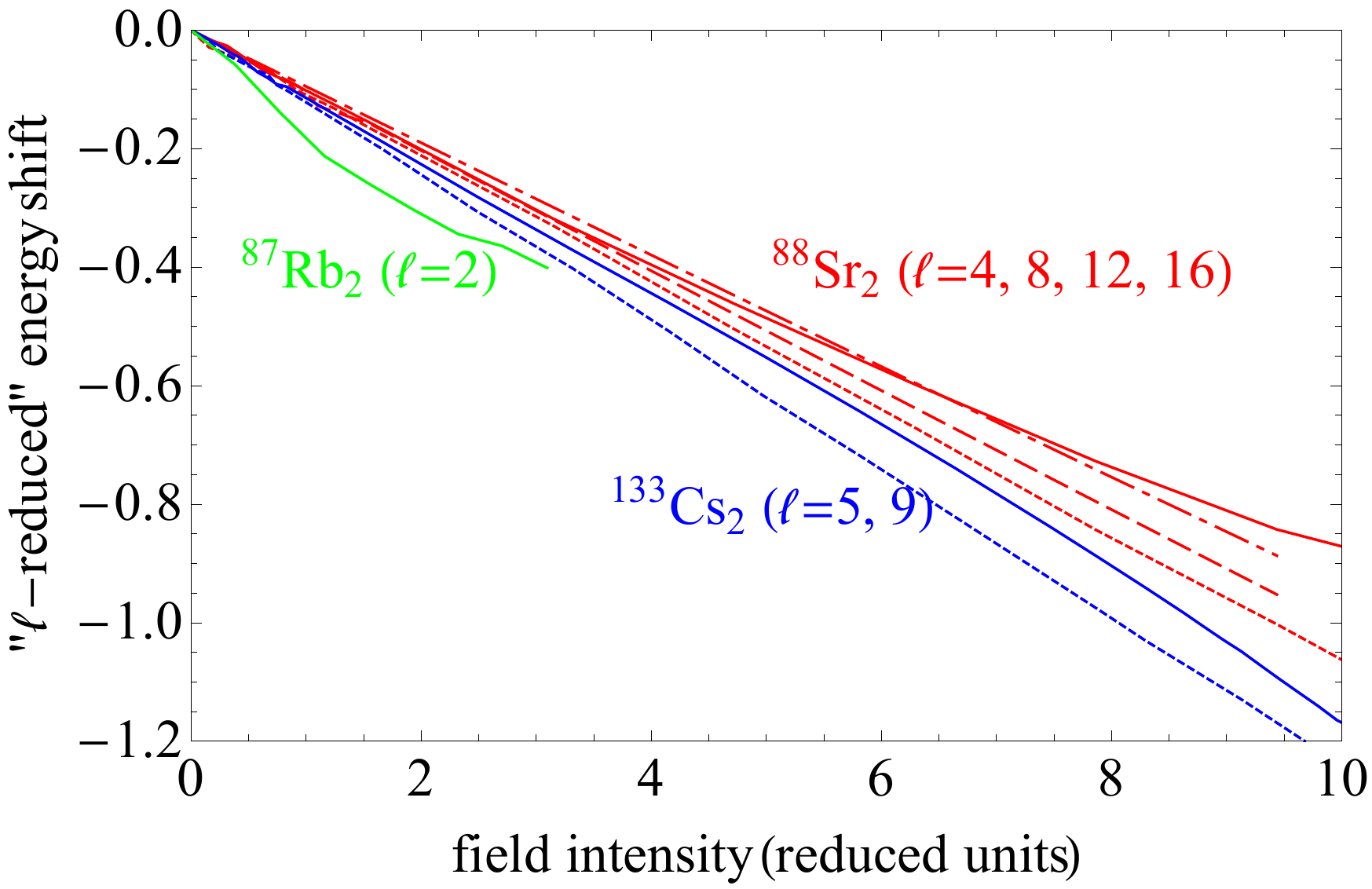}
  \caption{%(Color online) 
    Intensity dependence of the '$\ell$-reduced' 
    energy shift $\delta e/[\ell(\ell+1)]$, $\delta e=e_r^\ell(i)-
    e_r^\ell(0)$, i.e., the shift of the resonance position from its
    field free value, divided by $\ell(\ell+1)$.
    Shown are results obtained with the exact
    Hamiltonian~\eqref{eq:2D_Hamil}), converted to reduced units, 
    for $^{88}$Sr$_2$ (red, $\ell=4$: full line; $\ell=8$: dotted
    line; $\ell=12$: dashed line; $\ell=16$: dot-dashed  line), 
    $^{133}$Cs$_2$ (blue,  $\ell=5$: full line; $\ell=9$: dotted line)  
    and $^{87}$Rb$_2$ (green, $\ell=2$: full line.} 
  \label{fig:shift-reduit-rosario}
\end{figure}

To this end, we introduce '$\ell$-reduced' energy shifts or
slopes. These are the energy shifts or slopes, in reduced units, 
divided by $\ell(\ell+1)$. More precisely, for a field-dressed  
shape resonance adiabatically correlated to a field-free resonance in
the partial wave $\ell$, which occurs at the reduced energy $e_r^\ell(i)$
for a non-resonant field of reduced intensity $i$, the
$\ell$-reduced energy shift is equal to $\delta e/[\ell(\ell+1)]$,
where $\delta e=e_r^{\ell}(i)-e_r^{\ell}(0)$ denotes the shift of the
resonance from its field-free position (here, as everywhere else in
the paper, the position of a shape resonance is taken with  
respect to the field shifted dissociation limit). Analogously, 
we call the quantity $\delta e/[i\,\ell(\ell+1)]$ '$\ell$-reduced'
slope. The intensity dependence of the  $\ell$-reduced
energy shifts is reported in Fig.~\ref{fig:shift-reduit-rosario} for
the resonances in strontium, rubidium and cesium mentioned above. 
Except for the $\ell=2$ resonance in $^{87}$Rb$_2$, an almost linear
behavior is observed up to high intensity. Moreover, in the limit of
vanishing field all slopes are nearly equal.
%%%%%%%%%%%%%%%%%%%%%%%%%%%% 1 reduced energy shifts rosario %%%%%%%%%%%%%%
%%%%%%%%%%%%%%%%%%%%%%%%%%%%%%%%%%%%%%%%%%%%%%%%%%%%%%%%%%%%%%%%%%%%%%%%%%%

Before presenting a calculation of the slopes to first order
in the perturbation in Sec.~\ref{sec:pert}, 
we give here a simple qualitative argument, based on perturbation 
theory, to justify the approximate proportionality $\propto \ell(\ell+1)$ 
of the slope of the resonance position's intensity-dependence 
at vanishing intensity.  The interaction with the
field, given in reduced units by 
\begin{equation}
\label{eq:en-shift}
h_{int}=-i/x^3 \, (\cos^2 \theta -1/3)\,,
\end{equation}
can be treated as a perturbation and therefore, to 
first order, in a single-channel approach. One has thus to calculate
the matrix element of $h_f$ with the (single-channel) field-free wave
functions $f_\ell^{i=0}(x,\theta)$ 
(see Eq.~\eqref{eq:derideltapert1} in \ref{app:pert})
\begin{eqnarray}
\label{eq:el-matrix}
&&\left\langle f_\ell^{i=0}(x,\theta)\left|h_{int}
\right| f_\ell^{i=0}(x,\theta)\right\rangle =\nonumber \\
&&\quad\quad
-i\, \alpha(\ell) \int_{x_{0\ell}}^\infty\,[y_\ell^{i=0}]^2\, x^{-3}\,dx
\end{eqnarray}
with $m=0$ and 
\[
\alpha(\ell)=  \frac{2\ell(\ell+1)}{3(2\ell+3)(2\ell-1)}\,.
\]
A similar integral as the one in Eq.~\eqref{eq:el-matrix} occurs 
in the expectation value of $R^{-3}$ for field-free
wave functions obtained with the exact
Hamiltonian~\eqref{eq:2D_Hamil}, when $x_{0\ell}$ is replaced by 
$R=0$. It is important to note that the angular factor $\alpha(\ell)$
is nearly independent of $\ell$ and 
equal to $\sim 1/6$. The $\ell$-dependence of the matrix element,  
Eq.~\eqref{eq:el-matrix}, thus necessary arises from the radial part, i.e., the
expectation value of $x^{-3}$ for the field free wave functions. 
%As a result, we expect the energy shift with  
%respect to the field-shifted dissociation limit to   vary
%approximately as $\propto i\,\ell(\ell+1)$ in the limit of vanishing field.  

The field-free resonance positions result from the competition between the
van der Waals and centrifugal interactions at intermediate distance,
where the amplitude of the resonance wave function is very large. This
$x$-domain is located already in the asymptotic zone, but well before
the location $x_{\ell}$ of the potential barrier, i.e.,   
$x_{asym} < x_{0\ell} < x \ll  x_{\ell}$. Here $x_{asym}=R_{asym}/\sigma$
and $x_{\ell}$, $x_\ell= [\ell(\ell+1)]^{-1/4}$, is the point where the 
centrifugal and van der Waals interactions exactly balance each other
out. The amplitude of the $s$-wave ($\ell=0$) radial wave function,  
$y^{i=0}_{\ell=0}(x)$,  is  never resonant and thus always rather small in
this $x$-range. 
Close to threshold, the van der Waals interaction
prevails.  Treating the rotational kinetic energy as a perturbation
and accounting for the first order introduces a  resonant correction
proportional  to  $\ell(\ell+1)$ in the  $\ell$-wave radial
wavefunction of  the field-free molecule.
At intermediate distance $y^{i=0}_{\ell>0}(x) \approx~
y^{i=0}_{\ell=0}(x) + \ell(\ell+1) z(x)$. When evaluating 
the integral in  Eq.~\eqref{eq:el-matrix}, the zeroth order
contribution is small, and in the first order, only the cross term 
$y^{i=0}_{\ell=0}(x)\,  x^{-3}\, z(x) $ contributes
significantly. As the energy varies, resonant behavior
might appear, with a maximum of the radial integral  and therefore an energy shift with 
respect to the field-shifted dissociation limit approximately proportional 
to $i\,\alpha(\ell) \,\ell(\ell+1)$, i.e. proportional to  $i\,\ell(\ell+1)$.  

For non-zero non-resonant light, this result remains valid only as long as
the interaction with the light can be considered as a pertubation
compared to the centrifugal interaction, 
i.e., for $i\,\ll 6\, \ell(\ell+1)\,x_{0\ell}$, in the $x$-domain
mainly contributing to  the integral, i.e., for $x_{0\ell} < x < x_\ell$.
Therefore, for a given intensity $i$, the deviation of the reduced
energy shift from the approximate law is larger for smaller  
$\ell$-values.

The observation of an approximately universal intensity-dependence of
the reduced energy shifts can be equivalently formulated as follows:
In order to shift the position of two shape resonances with field-free 
rotational quantum numbers $\ell_1$ and $\ell_2$ in two molecules, 1
and 2, by the same amount, 
the reduced laser intensities $i_1$ and $i_2$ must be related as
\begin{equation}
  \label{eq:scale_i}
  i_1\ell_1(\ell_1+1)\approx i_2\ell_2(\ell_2+1)\,.  
\end{equation}
We emphasize that this rule  provides only the approximate
\textit{slope} of  
the energy variation at $i\rightarrow 0$. To obtain the energy variation 
itself, one also has to know the field-free reduced energy of the shape 
resonance, i.e., the $s$-wave scattering length~\cite{GaoPRA09,LondonoPRA10}.

%---------------------------------------------------------------------------
\section{Slopes at vanishing intensity from 
  perturbation theory}
\label{sec:pert}
Since the intensity dependence of the shape resonance positions
appears to be linear up to large values of the field
intensity, it is interesting  to study the behavior at very low
intensity and calculate the slopes at $i=0$.  
This procedure requires only free-field calculations, that is a
single-channel model. In principle, the 
perturbation theory has here to be applied to  a continuous 
spectrum.  We discuss the example of the $^{88}$Sr$_2$ shape
resonances  with field-free $\ell=4,~8,~12,~16$, whose positions and
widths  are recalled in table~\ref{tab:positionwidth88}. 
We present  results obtained by the single-channel nodal line technique,
with a detailed description of the resonance profiles, and compare
them to those obtained by solving the Schr\"odinger  
equation with Hamiltonian~\eqref{eq:2D_Hamil}, using a discretization
of the scattering continuum.  
%oooooooooooooooooooooooooooo 2 pos, width oooooooooooooooooooooooooooooooo
\begin{table}[tb]
  \centering
  \begin{tabular}{|l|c|c|c|c|c|}
    \hline 
    & $\ell=4$ & $\ell=8$ & $\ell=12$     & $\ell=16$          \\ \hline
    position (red. units)          & 33.26    & 139.3    & 247.2         & 253.5              \\ \hline
    position (mK)                  & 2.872    & 12.03    & 21.35         & 21.89              \\ \hline
    width (red. units)             & 11.48    & 0.9159   & 2.141~$10^{-3}$ & 8.778~$10^{-12}$ \\ \hline
    width ($\mu $K)                & 991.3    & 79.11    & 0.01849       & 7.582~$10^{-10}$   \\ \hline
  \end{tabular}
  \caption{Position and width of the $^{88}$Sr$_2$ field-free 
    shape resonances with $\ell$=4,~8,~12 and 16, both in reduced 
    and physical units, as calculated in paper
    I~\cite{crubellierNJP14} in a multi-channel asymptotic model 
    with the  nodal line technique.} 
  \label{tab:positionwidth88}
\end{table}
%ooooooooooooooooooooooooooooooooooooooooooooooooooooooooooooooooooooooooooo

%---------------------------------------------------------------------------
\subsection{Single-channel nodal line technique}
\label{subsec:continuum}
%---------------------------------------------------------------------------
The description of the pertubation of a shape resonance by a weak
interaction takes a rather simple form in the nodal line formalism. It
is described in Appendix~\ref{app:pert}, with no particular shape
of the potentials involved in the zero and first order expressions. 
In the case of interest here, the unperturbed asymptotic Schr\"odinger
equation for the radial 
wave function of wave $\ell$ reads, in reduced units, 
\begin{equation}
  \label{eq:schrodinger}
  \left[-\frac{d^2}{dx^2} - \frac{1}{x^6} + \frac{\ell(\ell+1)}{x^2}
    - e\right] y^{(0)} (x) = 0\,,
\end{equation}
where we have omitted, compared to Eq.~\eqref{eq:fct}, the index $\ell$
of the radial function $y(x)$ for simplicity. The superscript denotes
the order of perturbation theory. 
The perturbation is given by %equation Eq.~(\ref{eq:en-shift}), 
%more precisely by its average for a given $\ell$ value (see Eq.~\ref{eq:el_matrix})
%
\begin{equation}
\label{eq:en-shift2}
H_1=i\,v(x)=-i\,\frac{\alpha(\ell)}{x^3}\,.
%=-i~\frac{1}{x^3}\frac{2\ell(\ell+1)}{3(2\ell+3)(2\ell-1)} 
\end{equation}
Let us recall that in the nodal line  formalism (see Sec.~III~B of
paper~I~\cite{crubellierNJP14}), Eq.\eqref{eq:schrodinger} is only
solved in the asymptotic $x$ domain, $x>x_0$  
(here also the index $\ell$ is omitted for simplicity). The interactions
in the inner zone (potential, rotational kinetic energy) in the zeroth
order Hamiltonian are accounted for by the choice of the node
position, Eq.~\eqref{eq:nodal_line_i}.   
In the perturbative treatment, contributions coming from the inner
zone due to interaction with the non-resonant field, are
present in the first order of perturbation theory and have to 
be accounted for separately.

In a first step, we ignore the variation 
of the node position to treat the  problem in the outer zone. The
reduced slopes of position and width of the resonance  
can be obtained by calculating the perturbation of the energy profile
of the phaseshift %involved in the determination of the
                  %characteristics of 
characterizing the resonance structure  
(see Appendix~A~2 of paper~I~\cite{crubellierNJP14}). 
This perturbation, written in the Born
approximation~\cite{friedrich98}, cf. Eq.~\eqref{eq:derideltapert2}, 
is equal to  $\frac{d}{di}\Delta\delta_{out}(e)=-\pi\,\mathcal{I}_{out}(e)$
with 
\begin{eqnarray}
\label{eq:pert1}
\mathcal{I}_{out}(e)=-\frac{1}{\pi}\,\frac{d}{di}\Delta\delta_{out}(e)=
-\alpha(\ell) \int_{x_0}^{\infty}\frac{1}{x^3}[y^{(0)}(x)]^2 dx\,,
\end{eqnarray}
where $F_0(x)=y^{(0)}(x)$ is the field-free energy-normalized physical regular radial
function ($F_0(x_0)=0$). Assuming a Lorentzian shape of the derivative of the
%%% chr: Is there a good reason to assume a Lorentzian shape? If so,
%%% we should state it
phaseshift with respect to energy, one obtains for this derivative 
in Eq.~\eqref{eq:pert1} the right-hand side 
of Eq.~\eqref{eq:derideltap}. A simple fit procedure thus yields the
slopes at $i=0$ of both position and  
width of the resonance at $i=0$ (cf. first line of \autoref{tab:redslopespos88} 
and \autoref{tab:redslopesw88}), except for the extremely narrow resonance
with $\ell=16$, for which the slopes cannot be obtained. The simplified
formula of Eq.~\eqref{eq:formsimple} gives roughly the same result 
for the slope of the position as the fitting procedure (compare first
and third lines of \autoref{tab:redslopespos88}),  
except for the narrowest resonance ($\ell=4$).

The nodal line formalism allows us to also account for the modifications
due to the internal part of the perturbation which change the 
node position $x_0$. One can calculate the displacement of the
node positions with the complete Hamiltonian (see
%%% chr: what is meant complete Hamiltonian, full 2D (=exact) or asymptotic 2D?
paper~I~\cite{crubellierNJP14}): For instance, the 
slopes $\frac{dx_0}{di}$ of the intensity dependence of the node
positions at $i=0$ are listed in the last line of \autoref{tab:redslopespos88}
for $\ell=4,~8,~12,~16$ in $^{88}$Sr$_2$. If no reliable data are
available for the full potential, it is possible to use a 'universal'
value for these slopes (see Eq.~(13) in
paper~I~\cite{crubellierNJP14}). We show in the appendix that 
a simple relationship, Eq.~\eqref{eq:zi-2}, exists 
between the intensity dependence of the node
position and the corresponding modification of the slope 
of the intensity dependence of the phaseshift
$\frac{d}{di}\Delta\delta_{in}(e)=-\pi\,\mathcal{I}_{in}(e)$, 
 corresponding to the 
contribution of the inner zone, with
\begin{equation}
\label{eq:pert2}
\mathcal{I}_{in}(e)=-\frac{1}{\pi}\,\frac{d}{di}\Delta\delta_{in}(e)
=+\frac{dx_0}{di}\frac{1}{\pi^2\, G_0(x_0)^2}\,.
\end{equation}
voir signes 
Here $G_0(x_0)$ is the value at $x_0$ of the energy-normalized
irregular solution of the Schr\"odinger  
equation, which is orthogonal to the physical regular one, $F_0(x)=y^{(0)}(x)$,
with a node at $x_0$. Adding this 
quantity to the one coming from Eq.~\eqref{eq:pert1} and 
repeating the above fitting procedure for the sum
%, as above, to the second member of Eq.~\ref{eq:derideltap}, 
yields the total slopes of both  position and width of the resonance 
(see the second line in \autoref{tab:redslopespos88}
and \autoref{tab:redslopesw88}). 
The agreement with the slopes calculated  with the full potential is
excellent (compare lines 2, 4 and 5 of \autoref{tab:redslopespos88},
and lines 1 and 3 of \autoref{tab:redslopesw88}).  
%ooooooooooooooooooooooooooo 3 red slopes (position) ooooooooooooooooooooooo
\begin{table}[tb]
  \centering
  \begin{tabular}{|l|c|c|c|c|c|}
    \hline 
             & $\ell=4$  & $\ell=8$ & $\ell=12$ & $\ell=16$ \\ \hline\hline
 outer        & -0.0791   & -0.0710  & -0.0602   & -         \\ \hline
 outer+inner  & -0.0931   & -0.101   & -0.100    & -         \\ \hline \hline
 outer*       & -0.114    & -0.0713  & -0.0602   & -0.0521   \\ \hline 
 outer+inner* & -0.137    & -0.101   & -0.100    & -0.102    \\ \hline \hline
 calc cf I   & -0.0941   & -0.101   & -0.100    & -0.0954   \\ \hline\hline
 $dx_0/di$ &  4.417$\cdot 10^{-5}$   & 4.525$\cdot 10^{-5}$ &
 4.873$\cdot 10^{-5}$ & 5.497$\cdot 10^{-5}$  \\ \hline
  \end{tabular}
  \caption{Reduced slopes (i.e., slopes in reduced units, divided by
    $\ell(\ell+1)$) of the resonance position's intensity dependence
    for $^{88}$Sr$_2$ and $\ell$=4,~8,~12 and 16 
    from different approaches. 
    The first four lines are obtained with the single-channel
    perturbative approach described in the Appendix: Lines 1 and 3:
    including only the asymptotic outer part of the perturbation, Eq.~\eqref{eq:pert1}. 
    Lines 2 and 4: taking also the inner part through the
    intensity-dependence of the node position, Eq.~\eqref{eq:pert1},
    into account.  
    The first two lines correspond to a fitting procedure to formula
    Eq.~\eqref{eq:derideltap} of the derivative
    of the phase shift with respect to $e$ and $i$. The
    following two lines (marked by a star) correspond to the simplified
    formula Eq.~\eqref{eq:formsimple}, ignoring the slope of the
    intensity dependence of the 
    width. The fifth line results from a calculation similar to those  
    performed in  paper~I~\cite{crubellierNJP14}, i.e., with an
    asymptotic  single-channel model with $i$-dependent nodal lines.  
    The last line lists the 
    slopes of the node position's intensity-dependence 
    used in paper~I~\cite{crubellierNJP14} and required in lines~2
    and 4.} 
  \label{tab:redslopespos88}
\end{table}
%ooooooooooooooooooooooooooooooooooooooooooooooooooooooooooooooooooooooooooo
%ooooooooooooooooooooooooooo 4 red slopes (width)ooooooooooooooooooooooooooo
\begin{table}[tb]
  \centering
  \begin{tabular}{|l|c|c|c|c|c|}
    \hline 
    & $\ell=4$ & $\ell=8$ & $\ell=12$  & $\ell=16$   \\ \hline
    outer       & -1.44    & -0.213   & -0.0000891 & -           \\ \hline
    outer+inner & -1.88    & -0.324   & -0.000156  & -           \\ \hline 
    calc cf I  & -1.69    & -0.329   & -0.000160  & -           \\ \hline 
  \end{tabular}
  \caption{Same as \autoref{tab:redslopespos88} but for the
    width instead of the position of the resonances (lines 2 and 4
    are omitted since the simplified formula Eq.~\eqref{eq:formsimple}
    does not give any information on  
    the intensity dependence of the width).}
  \label{tab:redslopesw88}
\end{table}
%ooooooooooooooooooooooooooo 5 dx0/di oooooooooooooooooooooooooooooooooooooo 
% \begin{table}[tb]
%   \centering
%   \begin{tabular}{|l|c|c|c|c|c|}
%     \hline 
%     $\ell=4$       & $\ell=8$      & $\ell=12$     & $\ell=16$      \\ \hline

%   \end{tabular}
%   \caption{Slopes of the intensity dependence of the node position, $dx_0/di$, for the 
%     shape resonances with $\ell$=4,~8,~12 and 16 in $^{88}$Sr$_2$, 
%     as used also in paper~I~\cite{crubellierNJP14}.} 
%   \label{tab:dx0di}
% \end{table}
%ooooooooooooooooooooooooooooooooooooooooooooooooooooooooooooooooooooooooooo
%{\bf The text corresponding to Table VI is missing: Integration over energy along
%a lorentzian profile of the radial integral of $x^{-3}$ with  zero-order energy 
%normalized radial wave function (see Eq.\ref{eq:pert1}). For narrow resonance
%this area is equivalent to the 
%mean value of $x^{-3}$ for a discretized continuum  wave function at the resonance
%energy $e_r$ normalized to unit Limit for $\gamma \rightarrow 0$ of the integral
%of a lorentzion $\equiv \delta (e-e_r)$ . Results to be compared to Rosario value.
%But we note that the contribution for the inner zone is disregarded in our area.
%Cf the factor 1.75 obtained in I in the slopes for intensity indepent nodal lines}

%---------------------------------------------------------------------------
\subsection{Full potential calculations and discretized continuum}
\label{subsec:rosario}
%---------------------------------------------------------------------------
%oooooooooooooooooooooooooo 5 areas oooooooooooooooooooooooooooooooooooooooo
\begin{table}[tb]
  \centering
  \begin{tabular}{|l|c|c|c|c|c|c|}
    \hline 
    & $R_{max}$ [a$_0$]& $\ell=4$ & $\ell=8$ & $\ell=12$ & $\ell=16$   \\ \hline
    % outer       &   --   &  10.6    & 30.4     & 56.2      & 85.1    \\ \hline
    % outer + inner & --   &  12.6    & 44.5     & 97.3      & 159.3  \\ \hline \hline
    exact & $1\cdot 10^5$  & -0.1026         & -0.0756 %-0.0740
             & -0.1013         & -0.0952\\ \hline
    exact & $2\cdot 10^5$                             & -0.1010         &
    -0.0754         & -0.1017         & -0.0952\\ \hline
    PT  $\langle x^{-3}\rangle$ & $1\cdot 10^5$       & -0.137         &
    -0.0681         & -0.0900         & -0.0827\\ \hline
    PT $\langle x^{-3}\rangle$  & $2\cdot10^5$ & -0.133        &
    -0.0692         & -0.0904         & -0.0827\\ \hline
    full   PT     &          $1\cdot 10^5$           & -0.144         &
    -0.0743        & -0.1011         & -0.0952\\ \hline
    full PT & $2\cdot10^5$ & -0.140 &   -0.0756 & -0.1017 & -0.0952\\ \hline
  \end{tabular}
  \caption{%%
    %%% chr: I don't see why we should present the profile areas
    % Lines~1 and 2: Energy profile area, i.e., $\langle
    % x^{-3}\rangle$ (line~1: $-\mathcal{I}_{out}(e)/\alpha(\ell)$, 
    % Eq.~\eqref{eq:pert1}, line~2: $-\mathcal{I}_{in}(e)/\alpha(\ell)$, 
    % Eq.~\eqref{eq:pert2}) for the $^{88}$Sr$_2$ shape
    % resonances with $\ell$=4,~8,~12 and 16, 
    % calculated in a single channel for $i=0$.
    % in line 2 the value has
    % been corrected to include the contribution of the inner zone,
    % \textcolor[rgb]{1,0,0}{    as it is discussed in Sec.~(\ref{subsec:continuum}). 
    % Lines~3  to 8: 
    Reduced slopes  (i.e., slopes in reduced units, divided by $\ell(\ell+1)$)
    of the resonance position's intensity dependence for
    $^{88}$Sr$_2$, computed 
    from fitting a line to the results obtained with the exact
    Hamiltonian (lines~1 and 2) and computed within perturbation theory, 
    Eq.~\eqref{eq:energy_pt_x3} (lines~3, 4) and
    Eq.~\eqref{eq:energy_pt_full} (lines~5, 6) 
    for two different sizes $R_{max}$ of the discretization box. These
    slopes are to be compared to those reported in \autoref{tab:redslopespos88}.
  } 
  \label{tab:areas}
\end{table}
% The Schr\"odinger equation with the exact Hamiltonian~\eqref{eq:2D_Hamil}
% is solved by a hybrid computational
% approach, which combines a basis set expansion in terms of 
% associated Legendre functions for the angular coordinate and a mapped
% discrete variable representation for the internuclear distance with
% $R\le R_{max}$. 
%%% chr: this is already mentioned in Sec. II
The continuum spectrum obtained when representing the exact
Hamiltonian~\eqref{eq:2D_Hamil} in a box of size $R_{max}$
is discretized and represented
by means of $L^2$-normalized wave functions with energies $E_n>0$. 
Resonances in this discretized spectrum are identified by fitting the
rotational constant $\langle1/(2\mu R^{2})\rangle$ as a function
of energy  to a Lorentzian~\cite{LondonoPRA10}, where the expectation
value is computed using energy-normalized wave
functions~\cite{LucEPJD04,RosarioPRA07}. 
Very narrow resonances, such as those
with  $\ell=8$,
12 or 16 in $^{88}$Sr$_2$ (cf. \autoref{tab:positionwidth88} for their widths), appear
as $\delta$-functions at the 
corresponding resonance energy, independent of the size of the box
$R_{max}$. 
The intensity-dependence of the resonance positions
has been fitted to a line in the weak field regime,  $I \le
9\cdot 10^{6}$ W/cm$^2$ ($\sim 0.014$ in reduced units). The
corresponding slopes
are presented in \autoref{tab:areas} (lines 1 and 2), showing indeed
only a weak dependence on the  size of the discretization box. 

Analogously to the previous subsection, we treat 
the interaction with the non-resonant light, i.e., the last term 
in the full Hamiltonian \eqref{eq:2D_Hamil}, as a perturbation to the
field-free Hamiltonian, motivated by the approximately linear
dependence of the resonance positions on field intensity up to fairly
large intensity.
Time-independent perturbation theory provides the following
first-order correction to the field-free energy of the
resonances~\cite{FriedrichPRL95,RosarioNJP09}, 
\begin{eqnarray}
  \label{eq:energy_pt_full}
  \Delta E_{n,\ell} &=&
  -\frac{2\pi I}{c}\bigg[
  \langle \psi_{n,\ell}|\alpha_\perp(R)|\psi_{n,\ell}\rangle \nonumber\\
  &&+\langle \psi_{n,\ell}|\Delta\alpha(R)|\psi_{n,\ell}\rangle
  \frac{2\ell^2+2\ell-1}{(2\ell+3)(2\ell-1)}\bigg]\,,
\end{eqnarray}
where $\psi_{n,\ell}$ is the field-free  $L^2$-normalized wave
function of the resonance. This correction includes the shift of the
field-dressed dissociation limit, $E_0=-4\pi \alpha/c$.  
When using additionally the 
approximation~\eqref{eq:Hfield} for the
interaction Hamiltonian at sufficiently large distance,  the
first-order correction to the energy shift with respect to $E_0$, in
reduced  units, becomes 
\begin{equation}
  \label{eq:energy_pt_x3}
  \Delta e_{n\ell}=
  -i\langle \psi_{n,\ell}|x^{-3}|\psi_{n,\ell}\rangle \alpha(\ell)\,,
%  \frac{\ell(\ell+1)}{(2\ell+3)(2\ell-1)}\,.
\end{equation} 
equivalent to Eq.~\eqref{eq:el-matrix}.
Equations~\eqref{eq:energy_pt_full} and~\eqref{eq:energy_pt_x3} 
provide an approximation to the slopes of the intensity dependence of
the resonance positions, valid within the limits of a weak
perturbation. 
These slopes were computed for two different sizes of the
discretization box, 
using the field-free $L^2$-normalized wave functions, 
see~\autoref{tab:areas}. For $\ell=12$ and 16, 
the slopes obtained with perturbation
theory using the full interaction~\eqref{eq:energy_pt_full} (lines 5
and 6) are very close to the fit of the field-dressed energy versus
$i$ (lines 1 and 2), whereas the slopes obtained when 
neglecting the $R$-dependence of the polarizability anisotropy at
short range (lines 3 and 4) show some deviation.
This indicates the short-range part of the
interaction with the non-resonant field to be not completely
negligible. Comparing lines 1 and 2 in \autoref{tab:areas}
to \autoref{tab:redslopespos88} allows to assess the accuracy of the 
single-channel nodal line technique: For  $\ell=12$ and 16, the
error is below 10 per cent when using
intensity-dependent nodal lines (cf. lines 2 and 4 in
\autoref{tab:redslopespos88}). For $\ell=4$, the agreement between 
the full calculations and the single channel results is
slightly worse. This is not surprising since the intensity dependence
of this resonance is not perfectly linear even at low intensity,
cf. Fig.~3 of 
Ref.~\cite{GonzalezPRA12}. Thus this resonance cannot be fully
described by perturbation theory. This is evident both from comparing
lines 3, 4 and 5, 6 to lines 1, 2 in \autoref{tab:areas} and also from
comparing \autoref{tab:areas} and \autoref{tab:redslopespos88}.

\begin{figure}[tb]
  \centering
  \includegraphics[width=0.9\linewidth]{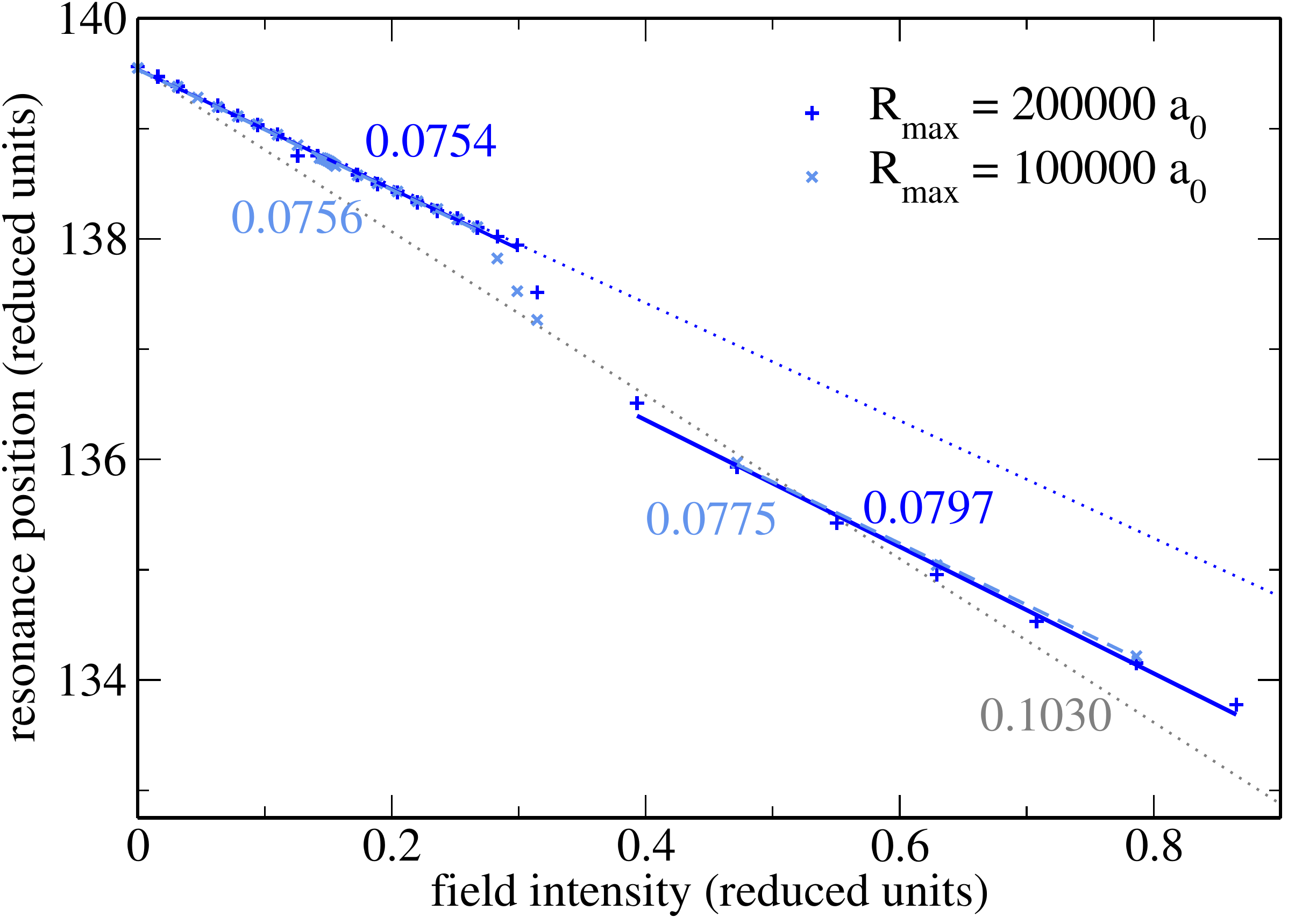}
  \caption{Intensity dependence of the position of the resonance with
    $\ell=8$ in $^{88}$Sr$_2$, obtained from calculations with the
    full potential for two different grid sizes (crosses). The lines
    represent linear fits to the data, with the numbers
    annotating the reduced slopes, i.e., the slopes divided by
    $\ell(\ell+1)$. The discontinuity is caused by an avoided crossing
    of the $\ell=8$ resonance and an over-the-barrier resonance with
    $\ell=6$ from the adjacent series.
    The grey dotted line is obtained when fitting the
    data over a large interval, ignoring the discontinuity due to the
    avoided crossing. } 
  \label{fig:res_l8}
\end{figure}
Finally, for $\ell=8$, the reduced slopes from the exact calculations
are surprisingly close to the less accurate single channel picture,
accounting only for the outer, but not the inner
region. In fact, the observed agreement of the slopes is accidental,
and perturbation theory breaks down in the case of $\ell=8$. This is
due to an avoided crossing occurs with an
$\ell=6$ resonance from an adjacent series which is over the barrier
at zero intensity, as illustrated in 
\autoref{fig:res_l8}.
The avoided crossing causes a discontinuity
around 3.5 reduced units in the otherwise linear intensity dependence
of the $\ell=8$ resonance. Since the resonance that comes close to
$\ell=8$ differs from $\ell$ by only 2, the wavefunctions are coupled
considerably even at very low non-resonant field
intensity. Consequently, 
right at the crossing, strong hybridization is
observed, with population distributed almost equally (44\% vs 56\%) in
the $\ell=6$ 
and $\ell=8$ channels. But for intensities just slightly away from the
avoided crossing, the mixing goes down to 10\% vs. 90\%. This explains the
perfectly linear intensity dependence that is resumed after the
avoided crossing is passed. 
When fitting the intensity dependence of
the resonance position for a range of
intensities which includes an avoided crossing, the slope is
overestimated. This is indicated by the grey dotted line in
\autoref{fig:res_l8}. Note that the behavior observed for the $\ell=8$
resonance is truly accidental, since it requires a
resonance with $\ell\pm 2$ from an adjacent series to come close by at
very weak field. Since this does not happen very often, perturbation theory
and the single channel approximation generally work very well, as
observed for $\ell=12$ and 16.

\section{Systematic single-channel calculations: 
  general trends}
\label{sec:single}

The good agreement between the perturbation theory results with exact
calculations shown in the previous section indicates that a
single-channel approximation may be sufficient for many
purposes. Specifically, we use the 
single-channel approximation in \autoref{sec:single-zero} 
to predict the position and
width of shape resonances (with no non-resonant field). This
extends the approach of Ref.~\cite{LondonoPRA10}, simplifying the
required calculations. Second, we employ the single-channel
approximation to investigate universality of the intensity-dependence
of shape resonances exposed to non-resonant light in
\autoref{sec:single-low-int}.

%---------------------------------------------------------------------------
\subsection{Field-free case}
\label{sec:single-zero}
%---------------------------------------------------------------------------
% The single channel calculations presented here aim 
% to complement the results of Ref.~\cite{LondonoPRA10} 
% (concerning $\ell=1$ to $\ell=6$ only), 
% which allow to estimate, from the knowledge of nodal lines (i.e. with known values 
% for $x_{00}$, $A$ and $B$ in Eq.~(\ref{eq:nodal_line_i}) and with $C=0$), the 
% position and width of any field-free shape resonance. 
%%%%%%%%%%%%%%%%%%%%%%%%%% 2 single channel 1 %%%%%%%%%%%%%%%%%%%%%%%%%%%%
\begin{figure}[tb]
  \centering
  \includegraphics[width=0.99\linewidth]{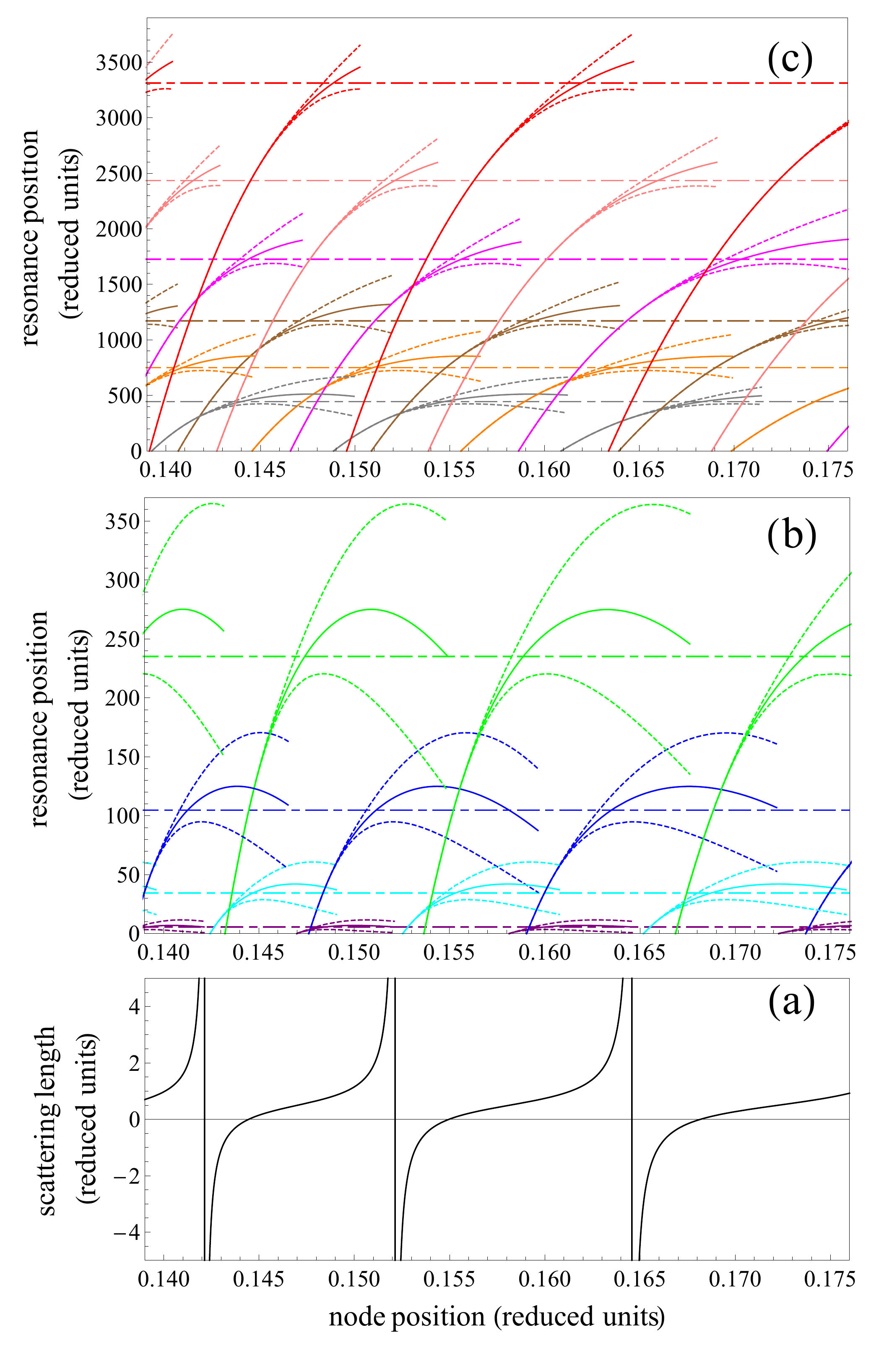}
  \caption{%(Color online) 
    b,c: Energies (taken with respect to the
    threshold) and widths of even $\ell$-wave %field-free 
    shape resonances as a function of the node  
    position $x_{0\ell}$ in a single-channel asymptotic model  
    with energy- and angular momentum-independent nodal lines. The
    energies, $e_r^\ell$, are represented by continuous lines; and the
    curves $e_r^\ell \pm \gamma^\ell/2$,  
    where $\gamma^l$ denotes the resonance width, are drawn as dashed lines. 
    The top of the potential barriers, $v_{top}^{\ell} =2
    [\ell(\ell+1)/3]^{3/2}$, are indicated by the horizontal
    dot-dashed lines (c: $10 \le \ell \le 20$, i.e., grey: $\ell=10$;
    orange: $\ell=12$; dark brown: $\ell=14$; light purple: $\ell=16$; 
    rose: $\ell=18$; red: $\ell=20$. 
    b:  $2\le \ell \le 8$: purple: $\ell=2$; light blue: $\ell=4$; 
    dark  blue: $\ell=6$;  green: $\ell=8$.) 
    a: $s$-wave scattering length, in reduced units.
  } 
  \label{fig:eliane}
\end{figure}
In the single-channel asymptotic model using the nodal line technique, the only 
parameter relevant for the partial wave $\ell$ is the node position, $x_{0\ell}$,
which determines the position and width of shape
resonances. The characteristics of the resonances can be determined
using the complex energy method, 
described Sec.~A~3 of paper~I~\cite{crubellierNJP14}, for narrow
resonances or employing the profile of the phase 
shift, see Sec.~A~2 in paper~I~\cite{crubellierNJP14}, for broad
resonances close to the top of the potential barrier. 
The resonance position as a function of 
node position is shown in Fig.~\ref{fig:eliane}
for even $\ell$ ranging from 2 to 20 and for node positions 
$0.139 \le x_{0\ell} \le
0.176$. In Fig.~\ref{fig:eliane}, a single
abscissa is used for the various $x_{0\ell}$, which is as if
the dependence of the nodal lines energy and angular momentum 
had been neglected, i.e., it is equivalent to $x_{00}=x_{0\ell}$,
or $A=B=0$ in  Eq.~\eqref{eq:nodal_line_i}.
 The resonance position
shows a pseudo-periodic dependence on node position: 
For each $\ell$ value, separate branches appear successively,
with discontinuities occuring at $\ell$-dependent values
$x_{0\ell}$ such that $J_{(2\ell+1)/4}(1/(2x_{0\ell}^2))=0$,  
where $J_\nu$ denotes a regular Bessel function. When this condition
is fulfilled, a
bound level with angular momentum $\ell$ reaches the threshold and
becomes a shape resonance. Correspondingly, the number of bound
$\ell$-levels is decreased by one, as is the 
number of nodes of the resonance wave function inside the potential 
barrier, $x_{0\ell}<x<x_{\ell}$. Between two consecutive discontinuities, 
along a given branch, the resonance energy increases from threshold,
reaches the top  of the potential barrier,
$v_{top}^{\ell}=2[\ell(\ell+1)/3]^{3/2}$, and even passes over the top
of the barrier, with the resonance profile becoming very broad. 
The dependence of the reduced $s$-wave scattering length on $x_{00}$
(see Eq.~(12) of Ref.~\cite{LondonoPRA10}) is also shown in 
Fig.~\ref{fig:eliane}. It exhibits an analogous pseudo-periodic
pattern, and 
the presence of an $\ell=0$ bound level just at threshold 
corresponds, as is well known, to infinite scattering length. 
One can see in Fig.~\ref{fig:eliane} that, for a given value of the 
scattering length, there is never more than one resonance 
below the top of a particular $\ell$-potential barrier; and  when 
a resonance appears in a particular $\ell$-channel, resonances also 
appear in the $\ell \pm 4\,p$ ($p$ integer) channels. 

%%%%%%%%%%%%%%%%%%%%%%%%%% 3 single channel 1 %%%%%%%%%%%%%%%%%%%%%%%%%%%%
\begin{figure}[tb]
  \centering
  \includegraphics[width=0.99\linewidth]{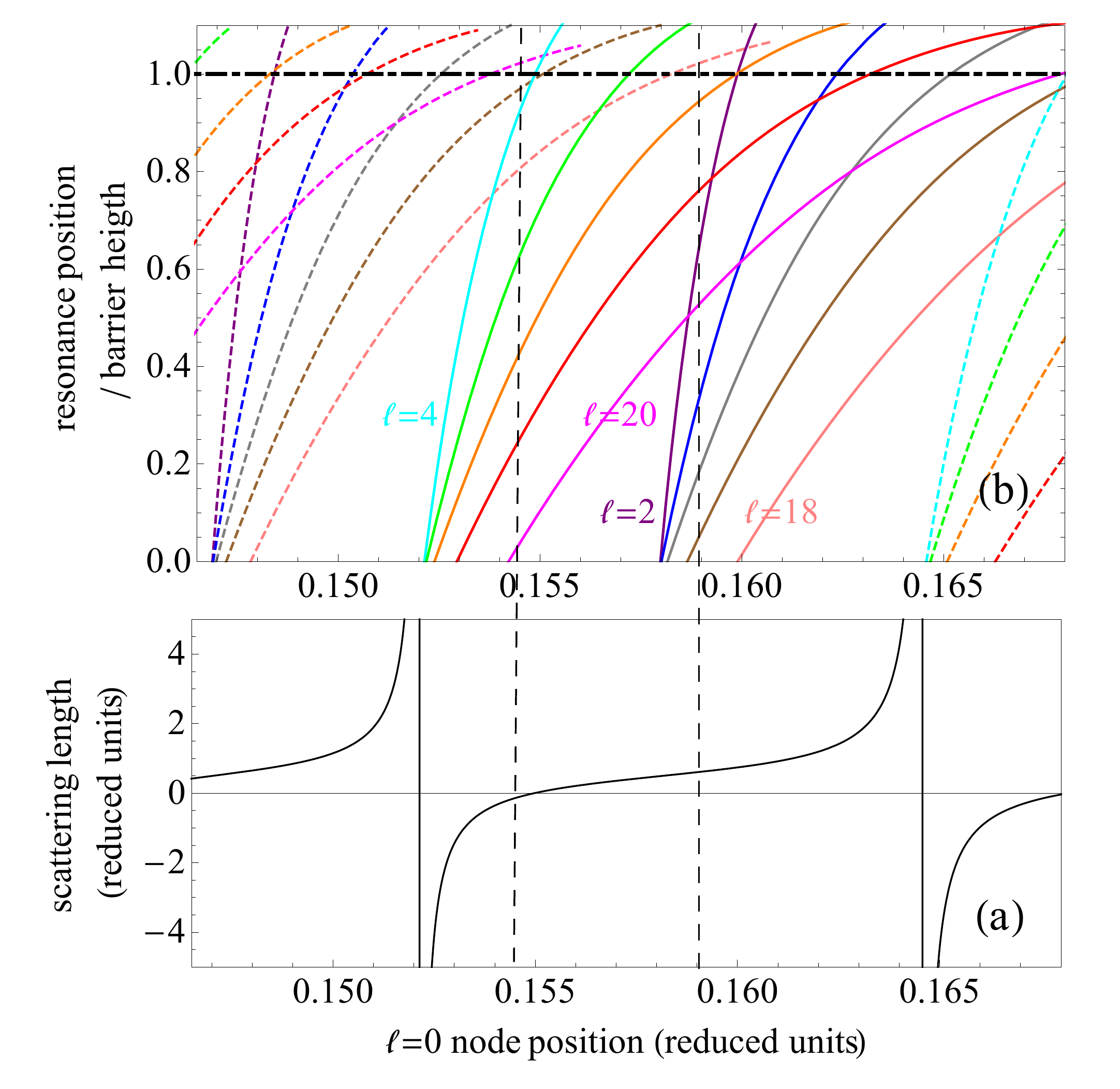}
  \caption{%(Color online) 
    b: Energy relative to the top of the potential
    barrier, $e_r^\ell/v^\ell_{top}$, of $\ell$-wave 
    shape resonances ($2 \le \ell \le 20$ even) as a function of the
    node position $x_{00}$ (in reduced units)
    of the $s$-wave threshold wave function ($e^{\ell=0}(x_{00})=0$) in the universal 
    single-channel asymptotic model with energy- and angular momentum-dependent nodal 
    lines ($A=A^G$ $B=B^G$). The relative energies for two
    $\ell+4p$-series (with integer $p$), $\ell=2,\ldots,18$
    and $\ell=4,\ldots,20$,  are represented by 
    continuous lines; and the three adjacent series by dashed lines
    (same color code as in Fig.~(\ref{fig:eliane}).  
    %%% chr: it is not clear what are the adjacent series
    The horizontal black dot-dashed line indicates the top of
    the potential barriers.  
    a: $s$-wave scattering length, $a_S(x_{00})$, as a function of
    node position.
    The vertical dashed black lines correspond to $^{88}$Sr$_2$ and 
    $^{86}$Sr$^{88}$Sr.}
  \label{fig:eliane:bo:gao}
\end{figure}
%%%%%%%%%%%%%%%%%%%%%%%%%%%%%%%%%%%%%%%%%%%%%%%%%%%%%%%%%%%%%%%%%%%%%%%%%%%
In order to properly account for the contribution of
the potential and  centrifugal term at short range, $x<x_{00}$, 
the nodal lines need to depend on both energy and angular
momentum~\cite{LondonoPRA10}. 
The results presented in Fig.~\ref{fig:eliane} can easily be
extended to energy- and angular momentum-dependent node positions
$x_{0\ell}$: The dependence on angular momentum simply introduces
$\ell$-dependent translations parallel to the horizontal axis.
The energy-dependence modifies the shape of the curves shown in 
Fig.~\ref{fig:eliane} only slightly. 
The position and width of field-free resonances 
of any molecule can be estimated from Fig.~\ref{fig:eliane}, once the nodal lines 
$x_{0\ell}$, Eq.~\eqref{eq:nodal_line_i} have been chosen
(we use in this section the 'universal' asymptotic model~\cite{LondonoPRA10},
with $A=A^G=-(x_{00})^7/8$, $B=B^G=(x_{00})^5/4$ and $C=0$,
resulting in nodal lines depending only on $x_{00}$). 
% %depending only on $x_{00}$ (see Sec.\ref{subsec:nodallines:multi}), i.e. 
% depending only on $x_{00}$, or, equivalently, on the $s$-wave
% scattering length. }
The results of the  transformation $e_r(x_{0\ell}) \rightarrow e_r(x_{00})$ are presented in
Fig.~\ref{fig:eliane:bo:gao}, 
with the abcissa now being actually $x_{00}$. In addition, for the figure 
to be more compact, we have divided $e_r^{\ell}$ 
by the height of the corresponding rotational barrier $v_{top}^\ell$. % The variation
% of the s-wave scattering length in reduced units is recalled in the same figure.
Figure~\ref{fig:eliane:bo:gao} is useful to predict, at least roughly,
the position of  shape resonances of a molecule with
$s$-wave scattering length $a_S$: The resonances
lie on the vertical line located at the abcissa 
$a(x_{00})=a_S$ (with $a_S$ in reduced units). The dashed vertical
lines in Fig.~\ref{fig:eliane:bo:gao} indicate the examples 
of $^{88}$Sr$_2$ and $^{86}$Sr$^{88}$Sr studied in paper~I, with
the scattering lengths  equal to $a_S=-2\,$a$_0$~\cite{SteinEPJD10}  
or -0.013 reduced units
and $a_S=97.9\,$a$_0$~\cite{ZhangCPL11} or
0.664 reduced units, respectively. Note that, for a
given molecule, the resonance 
energies relative to the barrier tops, $e_r^\ell/v_{top}^\ell$, generally 
decrease regularly with increasing $\ell$. 
Note also that shape resonances with 
$\ell=4p$ ($p$ integer, not too high) appear at threshold for a 
reduced scattering length with large absolute value, whereas 
shape resonances with $\ell=4p+2$ appear at threshold for a reduced scattering 
length close to 0.48. This property had been derived analytically by
Gao~\cite{GaoEPJD04} by solving the Schr\"odinger equation  for 
a $x^{-6}$ potential plus centrifugal term limited at $x_0^G\rightarrow 0$ by an infinite
repulsive wall~\cite{GaoEPJD04,MoritzPRA02}. With this potential,
analytical values for the reduced $s$-wave scattering length
$a^G_\ell$ and the wall position $x_{0\ell}^G$ for which the last,
least-bound rotational levels  
$\ell=1,\,2,\,3,\,4$ modulo $4$ are located exactly at the
dissociation limit, 
have been obtained~\cite{GaoEPJD04,MoritzPRA02}.
For $x_{00}\rightarrow 0$, the 'universal' 
asymptotic nodal line model becomes equivalent to Gao's universal model.

%%%%%%%%%%%%%%%%%%%%%%%%%%%%%% 4 single channel largeurs %%%%%%%%%%%%%%%%%%
\begin{figure}[tb]
  \centering
  \includegraphics[width=0.99\linewidth]{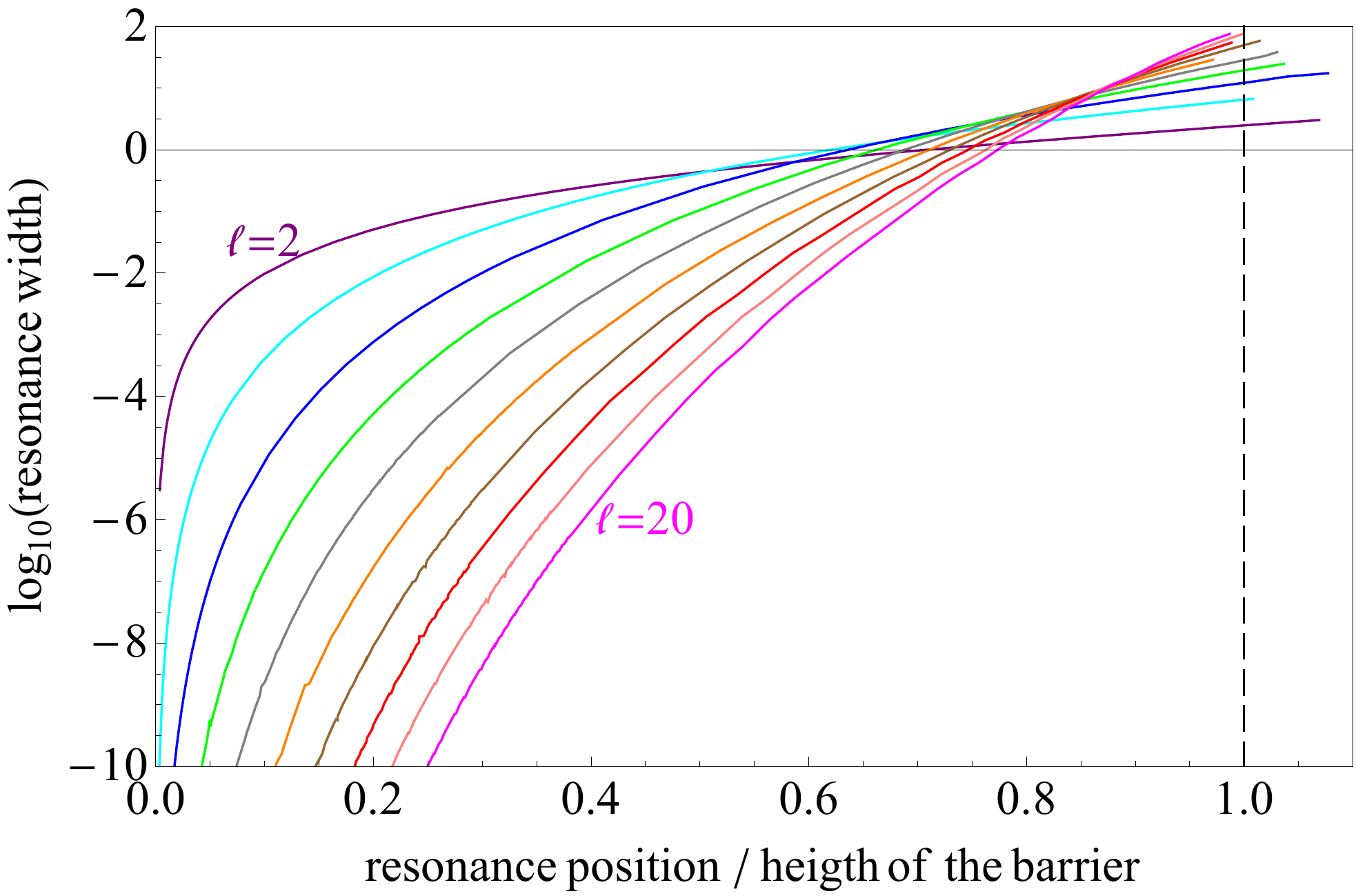}
  \caption{%(Color online) 
    Resonance widths $\gamma^\ell$ as a function
    of resonance position relative to the centrifugal barrier, 
    $e_r^\ell/v_{top}^\ell$, for  $\ell$-wave field-free 
    shape resonances (with $\ell$ even, $2 \le \ell \le 20$, same color
    code as in Fig.~\ref{fig:eliane}), obtained
    by single-channel calculations with energy- and angular 
    momentum-independent nodal lines. 
    The abscissas $0$ and $1$ correspond to a resonance at 
    threshold and a resonance at the top of the centrifugal barrier,
    respectively, and the 
    solid black horizontal line indicates the top of the barriers.} 
  \label{fig:eliane:gam}
\end{figure}
%%%%%%%%%%%%%%%%%%%%%%%%%%%%%%%%%%%%%%%%%%%%%%%%%%%%%%%%%%%%%%%%%%%%%%%%%%%
For completeness, Fig.~\ref{fig:eliane:gam} presents 
the widths of the field-free resonances,  calculated in the
single-channel asymptotic model, as a function of resonance energy. 
These widths are already visible in Fig~\ref{fig:eliane}, they are
represented by the distance between the 
dotted lines around each resonance.
%%% chr je ne comprends pas cette phrase:
% Their values in reduced units, $\gamma^\ell$, present a periodic 
% variation, the widths obtained for the same resonance energy $e_r^\ell$ but 
% calculated in different branches being not discernible. 
In contrast to Fig~\ref{fig:eliane}, Fig.~\ref{fig:eliane:gam}
emphasizes the general trend of 
the widths as a function of energy, relative to the top of
the barrier:
At threshold, the resonances have a vanishing width, which rapidly
increases when the resonance energy increases (note the logarithmic
scale). At the top of the potential barrier, the resonance width is
rather huge, between 1 to  100 reduced units. This general behavior is
observed for all partial waves. 
\subsection{Universality in the non-resonant light control of shape
  resonances at low intensity}
%Systematic single-channel calculations: low intensity field}
\label{sec:single-low-int}
%---------------------------------------------------------------------------

% In this section, we study systematically the influence, on shape resonances,
% of a non-resonant field at very low intensities, when the coupling between 
% the different channels is very small and we analyze the properties 
% of the dressed molecule  linear vs field 
% intensity. More precisely, we calculate the slopes at $i=0$ of the 
% evolution with $i$ of the field-dressed resonance position. 

The linearity of the intensity dependence of the resonance positions
observed in paper~I~\cite{crubellierNJP14} and the validity of
perturbation theory at low intensity suggest a more detailed
investigation of the field-dressed resonances in the universal
asymptotic model. To this aim, we determine the position of the
field-dressed resonances, $e_r^\ell(i)$,  
at very small field intensity ($i=0.01$ reduced units) as a function 
of node position $x_{0\ell}$, in the range $0.139 \le x_{0\ell} \le
0.176$, and for  partial waves with even $\ell$-value, with $2 \le
\ell \le 20$.  For the same partial wave and the same $x_{0\ell}$
value, the resonance position without non-resonant field is denoted by
$e_r^\ell(0)$.  When determining these two resonance positions,  
the contribution of the non-resonant field  
is accounted for only in the 'outer' zone, $x>x_{0\ell}$.
The contribution of the field to the $\ell$-reduced' 
slope of the resonance position's intensity-dependence 
in the outer zone can therefore  be
quantified as 
\begin{equation}
\label{eq:pente-out}
\mathcal{S}_{out}=\frac{e_r^\ell(i)-e_r^\ell(0)}{i\,\,l(l+1)}\,.
% {\mathrm {with}}\, i=0.01 \,{\mathrm{ reduced \, units}}\,.
\end{equation}
The contribution of the non-resonant field in the 'inner' zone
results, as described  in I~\cite{crubellierNJP14}, in a change of the
node position  proportional to the field intensity. With
Eq.~\eqref{eq:nodal_line_i}, this change  becomes 
$d\,x_{0\ell}/d\,i=C$. % The energy of the resonance
% depending on the node position (see Sec.~\ref{sec:single-zero} and
% Fig.~(\ref{fig:eliane})),
The contribution of the field to the $\ell$-reduced' slope in the
inner zone becomes 
\begin{equation}
\label{eq:pente-in}
%\frac{\Delta e}{\Delta i}{\huge{|}}_{in}=\frac{1}{\ell(\ell+1)} \frac{d\,e_r^\ell(0)}{d\,x_{0\ell}}\, \frac{d\,x_{0\ell}}{d\,i}\,.
\mathcal{S}_{in}=\frac{1}{\ell(\ell+1)} 
\frac{d\,e_r^\ell(0)}{d\,x_{0\ell}}\, \frac{d\,x_{0\ell}}{d\,i}\,,
\end{equation}
where the first derivative is evaluated from the depdendence of 
$e_r^\ell(0)$ on $x_{0\ell}$, reported in Fig.~\ref{fig:eliane}.
The second derivative is taken to be equal to  the value of $C$,
$C=C^G$, in the  universal asymptotic model (cf. Eq.~(13) in
paper~I~\cite{crubellierNJP14}),   
\begin{equation}
C^G=-x_{00}^4/12+3x_C^4/48 \,.
\label{eq:CG}
\end{equation}
Here, $x_C=R_C/\sigma$ is the position, in reduced units, at which 
the asymptotic expansion for the polarizabilities is truncated (see
Sec.~II of paper~I~\cite{crubellierNJP14}). 
In the universal model, the total reduced slope is equal to
$\mathcal{S}=\mathcal{S}_{in}+\mathcal{S}_{out}$ and 
depends for each $\ell$-value on the resonance position, $e_r^\ell(0)$.
Calculating $\mathcal{S}(e_r^\ell(0))$ for the same energy  
$e_r^\ell(0)$ but using  different branches for the node $x_{0\ell}$ 
(see Fig.~\ref{fig:eliane}) results in almost the same value.
% as it has been previously observed for the widths of the resonances (Fig.~(\ref{fig:eliane:gam})).
This proves the adequacy of our treatment of the different
interactions (potential, rotational 
energy, laser field interaction) in the inner zone.

%%%%%%% 5 single channel pentes-reduites %%%%%%%%%%%%%%%%%%
\begin{figure}[tb]
  \centering
	\includegraphics[width=0.99\linewidth]{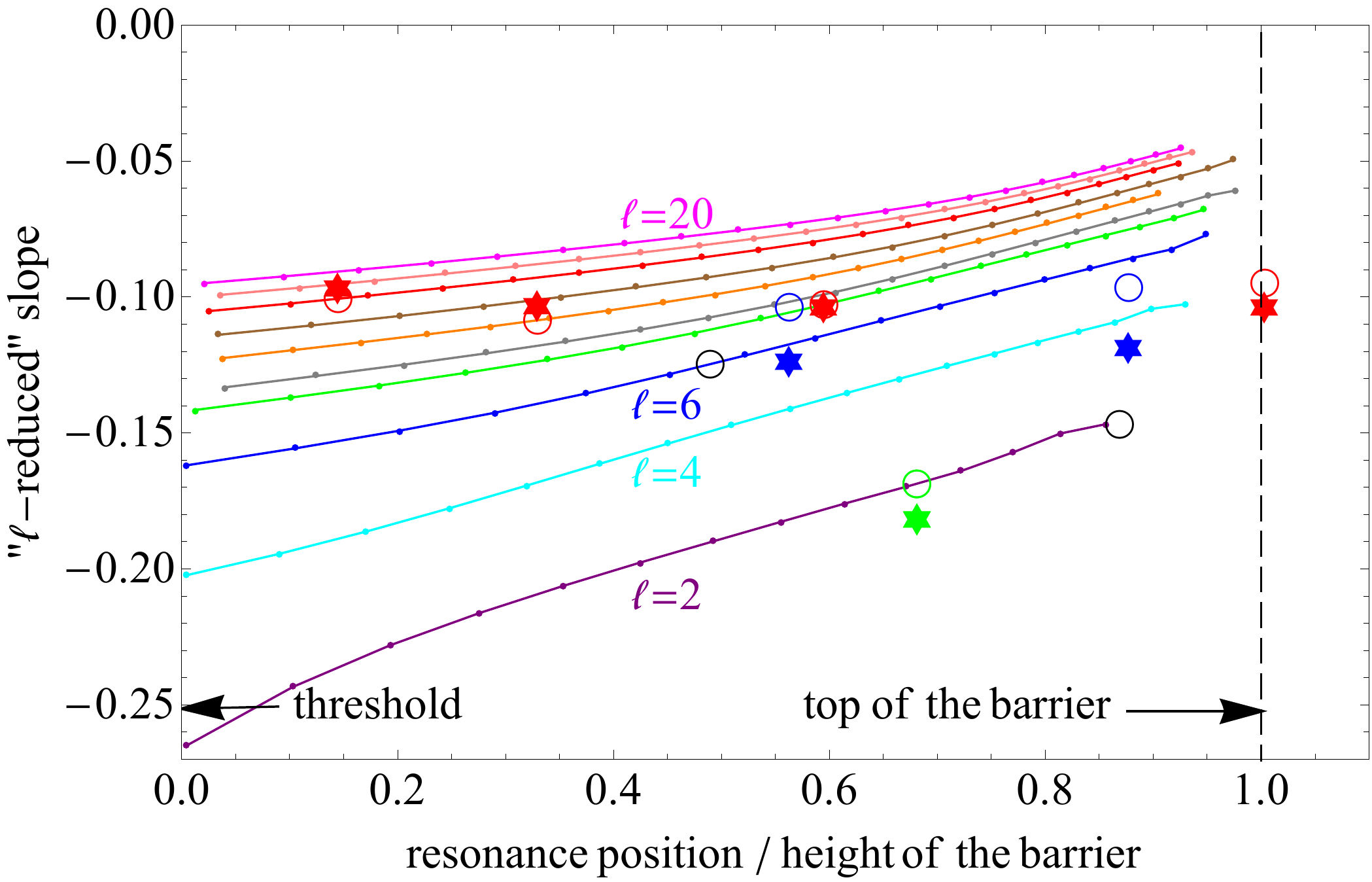}
  \caption{%(Color online)
    '$\ell$-reduced' slopes, $\delta e/i/[l(l+1)]$,  calculated with a
    single-channel universal  asymptotic model in the limit $i
    \rightarrow 0$,     as a function of $e_r^\ell(0)/v_{top}^\ell$ for 
    even $\ell$ values $2\le \ell \le 20$ (same color code as in
    Fig.~\ref{fig:eliane}).  
    The contribution of the interaction with the non-resonant
    light in both outer zone, $x>x_{0\ell}$, and inner zone,  $x<x_{0\ell}$,
    are accounted for, cf. Eqs.~\eqref{eq:pente-out} and~\eqref{eq:pente-in}. 
    The '$\ell$-reduced' slopes obtained by solving the multi-channel 
    Schr\"odinger equation for $^{88}$Sr$_2$ $\ell=4,~8,~12,~16$ 
    (red), $^{133}$Cs$_2$ $\ell=5,9$ (blue) and $^{87}$Rb$_2$ 
    $\ell=2$ (green), cf. Fig.~\ref{fig:shift-reduit-rosario},
    are indicated by stars.  The corresponding '$\ell$-reduced' slopes
    deduced from the universal asymptotic model  are indicated by open
    circles with the same colors as the stars. 
    The values predicted in paper I~\cite{crubellierNJP14} for the
    $\ell=2$ and $\ell=4$ resonances  in $^{86}$Sr$^{88}$Sr, using a
    multi-channel asymptotic model, are indicated by black open
    circles.
    } 
  \label{fig:eliane:slopes}
\end{figure}
%%%%%%%%%%%%%%%%%%%%%%%%%%%%%%%%%%%%%%%%%%%%%%%%%%%%%%%%%%%%%%%%%%%%%%%%%%%
Figure~\ref{fig:eliane:slopes} presents the total  $\ell$-reduced
slopes $\mathcal{S}$ as a function of 
the field-free resonance position relative to the 
heigth of the potential barrier, $e_r^\ell(0)/v_{top}^\ell$.
Although a large number of $\ell$ values and, in principle, 
any value of the scattering length is included in the calculations, 
strikingly, the slopes in Fig.~\ref{fig:eliane:slopes} are contained
in a rather small interval.  
The largest deviations occur for the lowest $\ell$-values, $\ell=2$ or 4.
For a fixed $\ell$-value, the slope increases slightly (i.e., its
absolute value decreases) with increasing resonance energy,   
or, equivalently, with increasing node position $x_{0\ell}$.
The absolute value of the '$\ell$-reduced' slope decreases 
with $\ell$. It presents a less pronounced variation when $\ell$
increases.  For $\ell$ larger than approximatively 8, the
'$\ell$-reduced' slope is roughly independent of $\ell$, with only a
weak dependence on the energy relative to the barrier height,
$e_r^\ell(0)/v^\ell_{top}$. The mean value of the total $\ell$-reduced
slopes is $\mathcal{S}(e_r^{\ell}=0)=-0.11\pm 0.2$ at the 
dissociation limit and $\mathcal{S}(e_r^{\ell}=v_{top}^\ell)=-0.05\pm
0.2$ at the top of the potential barrier. 

The $\ell$-reduced slopes obtained by solving the multi-channel 
Schr\"odinger equation with Hamiltonian~\eqref{eq:2D_Hamil} 
for the $\ell=4,\,8,\,12$ and 16 resonances of 
$^{88}$Sr$_{2}$,  the $\ell=2$  resonance of $^{87}$Rb$_2$ and the 
$\ell=5$ and 9 resonances of $^{133}$Cs$_2$, reported 
in Fig.~\ref{fig:shift-reduit-rosario},  are in very good agreement
with the values obtained from the  
universal asymptotic model, even for the highest $\ell$ values. The
universal model thus appears suitable to predict, at least
approximately, the $\ell$-reduced slope for any diatomic molecule.
Moreover, we find the heuristic scaling observed in
\autoref{sec:heuristic} to roughly hold in reduced units and for any 
value of the reduced scattering length. The approximate scaling rule
therefore seems to be generally applicable, for a large number of $\ell$ 
values and for any dimer.

%--------------------------------------------------------------------------
\section{Conclusions}
\label{sec:conclusion}
%--------------------------------------------------------------------------

We have applied first order perturbation theory to the asymptotic
model for shape resonance control of diatomic molecules interacting
with non-resonant light developed in Ref.~\cite{crubellierNJP14}. Our
work is the first to employ asymptotic model using the nodal line
technique~\cite{CrubellierEPJD99} to treat the perturbation of
continuum states in the framework of collision theory, i.e., the Born
approximation. As in earlier studies applying this approach to shape
resonances~\cite{LondonoPRA10,crubellierNJP14}, we find it crucial to
properly account for interactions at short range. 

The perturbation theory treatment had been 
motivated by observing a linear dependence of the resonance position
on non-resonant field intensity for several molecules with
different scattering lengths and shape resonances in different partial
waves. Comparison with full multi-channel
calculations has revealed our perturbative approach to be valid 
for not too low values $\ell$ of the partial waves. 
The advantage of the perturbative approach is to result in a
single-channel model which facilitates calculations
significantly. Although the non-resonant field couples  partial waves
with $\ell$ and $\ell\pm 2$, we find the single-channel perturbation approximation
to be  valid up to comparatively high intensity
We rationalize this
finding as follows: The first order perturbation 
correction to the resonance energy is related to the expectation value
of $x^{-3}$ (where $x$ denotes the interatomic separation in reduced
units).  The main contribution to the corresponding radial 
integral  comes from $x$ values just before the centrifugal barrier.
In this range, the amplitude of the scattering wave functions
is rather small, implying weak coupling, except at energies where a
resonance appears.  Since, close to threshold, resonances exist
simultaneously only for $\ell$ and  $\ell\pm 4$, the resonances
themselves are not coupled by the non-resonant field. Therefore an
overall only  weak channel mixing is observed. The only exception is a  
resonance with  $\ell\pm 2$ from an adjacent series coming close in
energy at very low field. In this case, the linear intensity
dependence is interrupted by discontinuities, indicating the
break-down of perturbation theory at the avoided crossing. However,
a single channel approximation is still able to correctly predict the
overall slope. That is, one only has to take care to determine the
slope from a sufficiently small intensity range where no avoided
crossing occurs.

We have analyzed the linear dependence of the resonance position
on non-resonant field intensity by introducing reduced slopes, where
the shift of the resonance position from its field-free value is
divided by $\ell(\ell+1)$. We have observed an almost identical value
for the reduced slope of several shape resonances in strontium,
rubidium and cesium. The approximately identical dependence of the
reduced slope on the energy, relative to the height of the centrifugal
barrier, is reproduced by systematic calculations using the universal
asymptotic model, where the field-free scattering length is the only
free parameter.
Our universal model is equivalent to the multi-channel quantum defect
treatment of shape resonances developed by
Gao~\cite{GaoJPB03,GaoEPJD04,GaoPRA09}. Fixing the value of the
field-free scattering length in the universal model 
allows for predicting the position of
field-free shape resonances~\cite{GaoPRA09,LondonoPRA10}. 
The corresponding predictions of our perturbative approach are more
accurate for lower partial waves. In contrast, the 
slopes of the intensity-dependence of the resonances are well
predicted even for high $\ell$-values.

For all partial waves except for $\ell=2$, 
the reduced slopes are found to vary regularly and in a small
intervall from the dissociation limit, where the resonances emerge, 
to the top of the centrifugal barrier, where the resonances start to
dissolve. This behavior is furthermore independent of the specific
molecule, it depends  neither on its reduced mass, 
nor on its $C_6$ coefficient, polarizability or 
scattering length, characteristic of the short range interaction.  
The stability of the reduced slope, derived here by
generalizing observations for a small number of molecules and partial
waves, presents a universal trend for field-dressed shape resonances.

The perturbative treatment developed here allows for a simple and
efficient approach to determine the intensity-dependence in
non-resonant light control of shape resonances since it requires
single-channel calculations using the field-free resonance functions
only. The slopes predicted by perturbation theory are sufficient to
estimate, at least approximately, the non-resonant field intensities
that are required to shift a field-free resonance to a desired
position. This is important for
utilizing non-resonant light control in molecule formation via
photoassociation~\cite{GonzalezPRA12} or Feshbach
resonances~\cite{TomzaPRL14}.
In addition to tuning the position and width of shape or
Feshbach resonances, non-resonant light control can also be employed
to change the $s$-wave scattering length. This will be studied in
detail elsewhere~\cite{AM_SL}.

%--------------------------------------------------------------------------
\begin{acknowledgments}
 Laboratoire Aim\'{e} Cotton is "Unit\'{e} Propre UPR 3321 du CNRS associ\'{e}e \`{a} 
l'Universit\'{e} Paris-Sud", member of the "F\'{e}d\'{e}ration Lumi\`{e}re Mati\`{e}re"
(LUMAT, FR2764) and of the "Institut Francilien de Recherche sur les Atomes Froids"
(IFRAF).
R.G.F. gratefully acknowledges a Mildred Dresselhaus award from the
excellence cluster "The Hamburg Center for Ultrafast Imaging
Structure, Dynamics and Control of Matter at the Atomic Scale" of the
Deutsche Forschungsgemeinschaft and financial support by the Spanish
Ministry of Science FIS2011-24540 (MICINN), grants P11-FQM-7276 and
FQM-4643 (Junta de Andaluc\'{\i}a), and by the Andalusian research
group FQM-207. 
\end{acknowledgments}
%--------------------------------------------------------------------------

%--------------------------------------------------------------------------
\appendix
%--------------------------------------------------------------------------
\section{Perturbation of a shape resonance in the 
  nodal line asymptotic model}
\label{app:pert}
%--------------------------------------------------------------------------
The description of the pertubation of a shape resonance by a weak
interaction takes a rather simple form in the nodal line
formalism. For simplicity, the following derivation assumes the use of
reduced units, but no particular forms {for the radial potentials involved in the 
zero order $H_0$ and first order  $H_1$ Hamiltonians.}
 
%shape of the potential $V_g(R)$.
We consider a shape resonance associated to a Hamiltonian $H_0$ 
with position $e_{r0}$ and width $\gamma_{0}$. Let us assume that the
resonance is characterized by a Lorentzian profile of the derivative 
$\delta_0 '(e)$ of the phaseshift with respect to energy, 
(see for instance Eq.~(1.185) in Ref.~\cite{friedrich98}), 
\begin{equation}
  \label{eq:zero}
  \delta_0 '(e) = \frac{\gamma_0/2}{(e-e_{r0})^2+ (\gamma_0/2)^2}\,. 
\end{equation}
In the nodal line asymptotic formalism, the energy-normalized radial wave 
function $y^{(0)}$ for any $\ell$ value 
at any value of scattering energy $e=k^2$ can be obtained from two separate
inward integrations in which the asymptotic behavior of the
energy-normalized solution is imposed to be either  
$\sin(kx)/\sqrt{\pi k}$ or $\cos(kx)/\sqrt{\pi k}$, with respective solutions 
$f_0(x)$ and $g_0(x)$ (see Sec.~Ã~2 in I). The condition imposed to the physical solution $y^{(0)}$ is 
to vanish at the node position $x=x_0$. The corresonding phaseshift is
given by 
\begin{equation}
  \label{eq:phaseshift}
  \tan[\delta_0(e))] = -\frac{g_0(x_0)}{f_0(x_0)}\,. 
\end{equation}
The  solution $y^{(0)}$ is identical to the regular wave 
function $F_0(x)$ associated to $H_0$, with an asymptotic
behavior $\sin(kx +\delta_0(e))/\sqrt{\pi k}$ (see
Ref.\cite{LondonoPRA10}), 
\begin{equation}
  \label{eq:regular}
  F_0(x)= \sin[\delta_0(e)]f_0(x)+\cos[\delta_0(e)]g_0(x)\,.
\end{equation}
The linearly independent solution associated to $F_0(x)$ is the irregular wave function 
$G_0(x)$  given by
\begin{equation}
  \label{eq:irregular}
  G_0(x)= +\cos[\delta_0(e)]f_0(x)-\sin(\delta_0(e)]g_0(x)\,. 
\end{equation}
%%%\textcolor[rgb]{1,0,0}{signe change pour avoir forma asymptotique $\cos(kx+\delta_0 )$ et signe correct A7}
%

Let us add a small perturbation characterized by a Hamiltonian $H_1=i \,v(x)$, 
where the parameter $i$ characterizes the strength of the perturbation. We 
assume the profile of the resonance to remain Lorentzian in the presence of 
this small perturbation, with the new profile depending on $i$, 
\begin{equation}
  \label{eq:non-zero}
  \delta '(e) = \frac{\gamma(i)/2}{(e-e_r (i))^2+ (\gamma(i)/2)^2}\,. 
\end{equation}
Close to $i=0$, the $i$-dependence of  $\delta '(e)$ is related to the 
derivatives {at $i=0$} of the $i$-dependencies of position and width: 
%{\bf j ai ajoute $\frac{1}{2}$ dans 
%$\frac{d\gamma}{di}$  et reference friedrich 98 dans shaperes-elk-2111}
%
\begin{eqnarray}
  \label{eq:derideltap}
  \frac{d}{di} \delta '(e) =
	&&\frac{de_r}{di}\frac{\gamma_0(e-e_{r0})}
	{\left[(e-e_{r0})^2+(\gamma_0/2)^2\right]^2}+\, \\ \nonumber
	&&\frac{d\gamma}{di}\frac{1}{2} \frac{(e-e_{r0})^2-(\gamma_0/2)^2}
	{\left[(e-e_{r0})^2+(\gamma_0/2)^2\right]^2}\,. 
\end{eqnarray}
The $i$-dependence of the phaseshift can be related to the Hamiltonian 
$H_1$: For any $e$, the additional phaseshift $\Delta \delta_{out}(e)$
coming from the  perturbation {in the outer asymptotic domain $x>x_0$}
 is given, to first order in $i$, by the Born approximation 
(see for instance Eq.~(4.38) in Ref.~\cite{friedrich98}):
\begin{eqnarray}
\label{eq:derideltapert1}
\Delta \delta_{out}(e)\sim\tan \Delta\delta_{out}(e)=-i\pi
\int_{x_0}^{\infty}v(x)[y^{(0)}(x)]^2 dx\,,
%\frac{d}{di}\delta(e)=-\pi \int_{x_0}^{\infty}v(x)y_0(x)^2 dx\,.
\end{eqnarray}
where $x_0$ is the position of the node, such that
\begin{eqnarray}
\label{eq:derideltapert2}
%\Delta \delta(e)\sim\tan \Delta\delta(e)=-i\pi \int_{x_0}^{\infty}v(x)y_0(x)^2 dx\, \\ \nonumber
\frac{d}{di}\Delta\delta_{out}(e)=-\pi \int_{x_0}^{\infty}v(x)[y^{(0)}(x)]^2
dx=-\pi\mathcal{I}_{out}(e)\,. 
\end{eqnarray}
The derivative with respect 
to $e$ of the last equation has to be fitted to 
Eq.~\eqref{eq:derideltap}, % \textcolor[rgb]{1,0,0}{reference changee} , 
allowing us to determine the 
derivatives with respect to intensity at $i=0$ of  
position and width of the field-dressed resonance. 

For narrow resonances, the $i$-dependence of the width is weak and the
corresponding term in $\frac{d\gamma}{di}$ can be neglected compared
to the other one. This implies that the integral in
Eq.~\eqref{eq:derideltapert2} has a Lorentzian shape, with the same  
center $e_{r0}$ and width $\gamma_0$ as $\delta_0'(e)$ (Eq.~\ref{eq:zero}) . If we call
$\mathcal{I}_{m}$ the  value  
of the integral $\mathcal{I}_{out}(e)$ at $e=e_{r0}$,  we deduce the slope of the $i$-dependence  
of the resonance position at $i=0$ to be 
\begin{equation}
\label{eq:formsimple}
\frac{d}{di}e_r=+\pi \mathcal{I}_{m} \gamma /2 =\int_{-\infty}^{\infty}\mathcal{I}_{out}(e)\, de\,,
\end{equation}
which is equal to the strength of the interaction integrated over
the whole energy-profile. 
For an attractive potential $v(x)<0$, the integral 
$\mathcal{I}(e)$ and therefore the slopes $\frac{de_r}{di}$ are negative and 
$\Delta_{out}(e)$, Eq.~\eqref{eq:derideltapert1}, is positive.

The nodal line formalism also allows to account for the perturbation due to
the internal part of the perturbation  $H_1$ at $x<x_0$ , which %is taken into account
%by the choice of the   node position. Indeed the perturbation $v(x)\, i$
 introduces a shift $\Delta x$ proportional to $i$ in the node position.
A simple relationship between $\Delta x$
%the dependence of the node position vs $i$, $\frac{dx_0}{di}$,  
and the corresponding modification to first order in $i$ of the phaseshift 
$\Delta \delta_{in}(e)$ can be obtained
from Eq.(\ref{eq:phaseshift}) 
\begin{equation}
\label{eq:zi-1}
\frac{d}{dx_0}\Delta\delta_{in}(e)=-\frac{W(g_0,f_0)}{f_0(x_0)^2+g_0(x_0)^2}\,,
\end{equation}
where $W(g_0,f_0)=g_0\,f'_0 \,-\, g'_0\,f_0$ denotes the Wronskian (here the derivatives
are taken with respect to $x$).
Employing the property $F_0(x_0)=0$ and the relation between the pairs of functions 
$(f_0,g_0)$ and $(F_0,G_0)$,
\begin{eqnarray}
\label{eq:reg-irreg}
W(g_0,f_0)&&=W(G_0,F_0)=\frac{1}{\pi} \\ \nonumber
f_0(x_0)^2+g_0(x_0)^2&&=F_0(x_0)^2+G_0(x_0)^2=G_0(x_0)^2 \,,
\end{eqnarray}
 we find 
\begin{equation}
\label{eq:zi-2}
\frac{d}{dx_0}\Delta \delta_{in}(e)=-\frac{1}{\pi\, G_0(x_0)^2}\,.
\end{equation}
Finally, the contribution of the inner zone to the variation 
of the phaseshift is, for any value of energy,
\begin{equation}
\label{eq:zit-2}
\frac{d}{di}\Delta \delta_{in}(e)=-\frac{dx_0}{di}\frac{1}{\pi\, G_0(x_0)^2}\,.
\end{equation}
For an attractive potential $v(x)$, the shift in the node position is
negative (see the value of $C^G$ in Eq.~(13) 
in paper I~\cite{crubellierNJP14}) and $\Delta \delta_{in}(e)$ is positive.

This additional correction to the phaseshift coming from the   the inner part of the perturbation
also results in a shift of the resonance position. 
This contribution Eq.~(\ref{eq:zit-2}) is to be  added to the slope due to the asymptotic 
part of the perturbation Eq.~(\ref{eq:derideltapert2}).
As above the derivative with respect to $e$ of the total slope of the change in the phaseshift
associated with the perturbation $H_1$
\begin{equation}
\label{eq:zit-tot}
\frac{d}{di}\Delta \delta(e)=\frac{d}{di}\Delta \delta_{out}(e) + \frac{d}{di}\Delta \delta_{in}(e)=-\pi \mathcal{I}(e)\,.
\end{equation}
has to be fitted to Eq.(\ref{eq:derideltap}) to determine the slopes at $i=0$ in the variation 
of the energy position and width of the field-dressed resonances.
%This has to be added to the similar term in the second line of 
%Eq.~\eqref{eq:derideltap}. It allows for deducing the total value of
%the slope of the $i$-dependence of the resonance position and,
%eventually (if the resonance  is not too narrow), of the resonance
%width, by the same fitting procedure as described for
%Eq.~\eqref{eq:derideltapert2}.

%%%%%%%%%%%%%%%%%%%%%%%%%%%%%%%%%%%%%%%%%%%%%%%%%%%%%%%%%%%%%%%%%%%%%%%%%%
%\bibliography{shaperes2}
%%%%%%%%%%%%%%%%%%%%%%%%%%%%%%%%%%%%%%%%%%%%%%%%%%%%%%%%%%%%%%%%%%%%%%%%%%

\end{document}